\documentclass[11pt]{article}
\usepackage{graphicx}
\usepackage{amsmath}
\usepackage{mathptmx}
\usepackage[pdftex]{hyperref}
\usepackage{amssymb,latexsym,fancyhdr,color,graphics,graphicx}

\newtheorem{theorem}{\bf Theorem}[section]

\newtheorem{remark}{\bf Remark}[section]
\newtheorem{acknowledgements}{\bf Acknowledgements}[section]
\newtheorem{prof}{\bf Prof}[section]

\newcommand{\bq}{\begin{equation}}
\newcommand{\eq}{\end{equation}}
\newcommand{\be}{\begin{eqnarray*}}
\newcommand{\ee}{\end{eqnarray*}}
\newcommand{\ben}{\begin{eqnarray}}
\newcommand{\een}{\end{eqnarray}}

\begin{document}
%\today
\renewcommand{\theequation}{\thesection.\arabic{equation}}
\begin{center}
{\Large \bf Optimization problem for a portfolio with an illiquid asset: Lie group analysis}
\\[2ex]
 { \large Ljudmila A. Bordag, Ivan P. Yamshchikov\\
Faculty of Mathematics and Natural Sciences,\\ University of Applied Sciences Zittau/G\"orlitz,
Theodor-K\"orner-Allee 16,\\
D-02763 Zittau, Germany
}
\end{center}

\begin{abstract}
Management of a portfolio that includes an illiquid asset is an important problem of modern mathematical finance. One of the ways to model illiquidity among others is to build an optimization problem and assume that one of the assets in a portfolio can not be sold until a certain finite, infinite or random moment of time. This approach arises a certain amount of models that are actively studied at the moment.

Working in the Merton's optimal consumption framework with continuous time we consider an optimization problem for a portfolio with an illiquid, a risky and a risk-free asset. Our goal in this paper is to carry out a complete Lie group analysis of PDEs describing value function and investment and consumption strategies for a portfolio with an illiquid asset that is sold in an exogenous random moment of time with a prescribed liquidation time distribution. The problem of such type leads to three dimensional nonlinear Hamilton-Jacobi-Bellman (HJB) equations. Such equations are not only tedious for analytical methods but are also quite challenging form a numeric point of view. To reduce the three-dimensional problem to a two-dimensional one or even to an ODE one usually uses some substitutions, yet the methods used to find such substitutions are rarely discussed by the authors.

We find the admitted Lie algebra for a broad class of liquidation time distributions in cases of HARA and log utility functions and formulate corresponding theorems for all these cases.
We use found Lie algebras to obtain reductions of the studied equations. Several of similar substitutions were used in other papers before whereas others are new to our knowledge. This method gives us the possibility to provide a complete set of non-equivalent substitutions and reduced equations.

\end{abstract}
%\keywords{portfolio optimization \and illiquidity \and viscosity solutions \and random income}

\section{Introduction}
Study of optimization problems with an illiquid asset leads to three dimensional nonlinear Hamilton-Jacobi-Bellman (HJB) equations. Such equations are not only tedious for analytical methods but are also quite challenging form a numeric point of view. One of the standard techniques is to find an inner symmetry of the equation and reduce the number of variables at least to two or if possible to one. The problems of lower dimensions are usually better studied and are, therefore, easier to handle.

All papers known to us devoted to three dimensional HJB equations provide variable substitutions without any remark on how to get similar substitution in other cases or why they use this or that substitution.
Yet since the famous work of S. Lie \cite{Lie1891} it is well known that smooth point transformations with continuous parameter admitted by linear or nonlinear partial differential equations (PDEs) can be found algorithmically using Lie group analysis. The procedure that helps to find a symmetry group admitted by a PDE is well described in many textbooks, for example, we can recommend \cite{Olver}, \cite{Ibragimov1994} or \cite{Bordag2015} to the interested reader. Yet practical application of these procedures is connected with tedious and voluminous calculations which can be only slightly facilitated with the help of modern computer packages. For example, preparing this paper we used the program {\bf IntroToSymmetry} to obtain the determining system of equations. Solving these determining systems of partial differential equations is usually rather hard and can rarely be done by an algorithm, but the possibility to find the system of determining equations facilitates the work of a researcher since the systems are quite voluminous. For example, in the studied cases the systems had more then a hundred equations.

Once the Lie algebra admitted by the studied PDE is found one can find all non equivalent variable substitutions which reduce the dimension of the given PDE, if there are any. The found Lie algebra admitted by the PDE generates the corresponding symmetry group of this equation. Using the corresponding exponential map of the adjoint representation of the considered Lie algebra we can find the symmetry group or corresponding subgroups of the equation as well. We do not have to look for an explicit form of the symmetry group to find reductions of the studied PDEs and invariant solutions of the equations. It is enough to know and to use the properties of the symmetry algebra which corresponds to the admitted symmetry group. The optimal system of subalgebras of this algebra gives rise to a complete set of non equivalent substitutions and as a result a set of different reductions of the studied  PDE.

The solutions of reduced PDEs are called invariant solutions because they are invariant under the action of a given subgroup. The goal of this paper is to find the admitted Lie algebras for PDEs describing value function and investment and consumption strategies for a portfolio with an illiquid asset that is sold in a random moment of time with a prescribed liquidation time distribution. We find the admitted Lie algebras for a certain class of liquidation time distributions in cases of HARA and log utility functions and formulate  corresponding theorems. We provide the optimal system of subalgebras for a general case of a liquidation time distribution in both cases of HARA and logarithmic utility functions.  We separately regard a case of an exponential distribution of a liquidation time where the corresponding PDE admits an extended Lie algebra. It leads to certain distinguishing properties that give rise to non trivial reductions of three dimensional PDEs to two dimensional equations and even to ordinary differential equations in some cases. We describe all non equivalent substitutions, provide the reductions and the corresponding lower dimensional equations as well as the corresponding allocation-consumption strategies.

\section{Economical setting} \label{problem}

In \cite{boyazh}, \cite{BordagYamshchikovZhelezov} the authors described in detail an optimization problem that corresponds to the following situation: an investor has an illiquid asset that has some paper value and can not be sold till some moment of time that is random with a prescribed distribution. He tries to maximize his average consumption investing into a risky asset that is partly correlated with the illiquid one and into a riskless asset, that has a constant dividend rate.

\subsection{Formulation of the optimization problem}

The investor's portfolio includes a riskless bond $B_t$, a risky asset $S_t$ and a non-traded asset $H_t$ that generates stochastic income, i.e., dividends or costs of maintaining the asset. The liquidation time  of the portfolio $T$ is a randomly-distributed continuous variable.
The risk-free bank account $B_t$, with the interest rate $r$, follows
	
	\begin{equation} \label{bond_r}
		dB_t = rB_t\,dt, \, t \leq T,
	\end{equation}
where $r$ is constant.
 The stock price $S_t$ follows the geometrical Brownian motion
	
	\begin{equation} \label{asset_S}
		dS_t = S_t(\alpha\, dt + \sigma\, dW^{1}), \, t \leq T,
	\end{equation}
with the continuously compounded rate of return $\alpha > r$ and the standard deviation $\sigma$. The lower case index $t$ denotes the spot value of the asset at the moment $t$.
The illiquid asset $H_t$, that can not be traded up to the time $T$ and its paper value is correlated with the stock price and is governed by

	\begin{equation}
		\frac{dH_t}{H_t} = (\mu - \delta)\,dt + \eta(\rho\, dW^{1} + \sqrt{1 - \rho^2}\,dW^{2}_t), \, t \le T,
	\label{eq:H1}
	\end{equation}
as it was shown in \cite{BordagYamshchikovZhelezov}, where $\mu$ is the expected rate of return of the risky illiquid asset, $(W^{1}, W^{2})$ are two independent standard Brownian motions, $\delta$ is the rate of dividend paid by the illiquid asset, $\eta$ is the standard deviation of the rate of return, and $\rho \in(-1, 1)$ is the correlation coefficient between the stock index and the illiquid risky asset. The parameters $\mu$,  $\delta$, $\eta$, $\rho$ are all assumed to be constant.\\
	
The stochastically distributed time $T$ is an exogenous time and it does not depend on the Brownian motions  $(W^{1}, W^{2})$. The probability density function of the $T$ distribution is denoted by $\phi(t)$, whereas $\Phi(t)$ denotes the cumulative distribution function, and  $\overline{\Phi}(t)$, the survival function, also known as {\em a reliability function}, $\overline{\Phi}(t) = 1 - \Phi (t)$. We omit here the explicit notion of the possible parameters of the distribution in order to make the formulas shorter.\\

We assume that the investor consumes at rate $c(t)$ from the liquid wealth and the allocation-consumption plan $(\pi, c)$ consists of the allocation of the portfolio with the cash amount $\pi = \pi(t)$ invested in stocks, the consumption stream $c = c(t)$ and the rest of the capital kept in bonds. Further on we sometimes omit the dependence on $t$ in some of the equations for the sake of clarity of the formulas.  The consumption stream $c$ is admissible if and only if it is positive and there exists a strategy that finances it. All the income is derived from the capital gains and the investor must be solvent. In other words, the liquid wealth process $L_t$ must cover the consumption stream.
The wealth process $L_t$ is the sum of cash holdings in bonds, stocks and {\em random} dividends from the non-traded asset minus the consumption stream, i.e. it  must satisfy the balance equation

	\begin{eqnarray}
		 dL_t &=& \bigl(rL_t + \delta H_t + \pi_t(\alpha - r) - c_t\bigr)\, dt + \pi_t\sigma\, dW^1. \nonumber
	\end{eqnarray}

The investor wants to maximize the overall utility consumed up to the random time of liquidation $T$, given by

	\begin{equation}
		\mathcal{U}(c) := E \left[\int^\infty_0{\overline{\Phi}(t)U(c)}\,dt\right]. \label{eq:supinfty}
	\end{equation}

It means we work with the problem (\ref{eq:supinfty}) that corresponds to the \emph{value function} $V(l,h,t)$, which is defined as

	\begin{equation} \label{valueFun}
		V(l, h, t) = \max_{(\pi, c) } E \left[ \int_t^\infty \overline{\Phi}(t)U(c)\, dt \;|\; L(t) = l, H(t) = h \right],
	\end{equation}
where $l$ could be regarded as an initial capital and $h$ as a paper value of the illiquid asset. The value function $V(l, h, t)$ satisfies  the Hamilton--Jacobi--Bellman (HJB) equation for the value function, in terms of  $l$, $h$ and $t$

	\begin{eqnarray}
		V_t (l, h, t) &+& \frac{1}{2}\eta^2h^2\,V_{hh} (l, h, t) + (r l + \delta h)\,V_l (l, h, t) + \nonumber \\
  		& & (\mu - \delta) h\,V_h (l, h, t) + \max_{\pi} G[\pi] + \max_{c \geq 0} H[c]\;\; =\;\; 0, \label{eq:HJB21}\\
		G[\pi] &=& \frac{1}{2}V_{ll}(l, h, t)\,\pi^2 \sigma^2 + V_{hl}(l, h, t)\,\eta\rho\pi\sigma h %\nonumber \\
      		+ \pi(\alpha - r)\,V_l(l, h, t), \label{eq:Gmax21} \\
		H[c] &=& -c\,V_l (l, h, t) + \overline{\Phi}(t) U(c), \label{eq:Hmax21}
	\end{eqnarray}

with the boundary condition

	\begin{equation} \label{cond:main}
		V(l, h, t) \to 0, ~~ as ~~ t \to \infty.
	\end{equation}

In \cite{boyazh} and \cite{BordagYamshchikovZhelezov} the authors have already demonstrated that the formulated problem has a unique solution under certain conditions. Namely,
\begin{enumerate}
\item $U(c)$ is strictly increasing, concave and twice differentiable in $c$,
\item $U(c) \leq M(1+c)^{\gamma}$ with $0<\gamma<1$ and $M>0$,
\item $\lim_{c \to 0} U'(c) = + \infty$,  $\lim_{c \to + \infty} U'(c) = 0$.
\item $\lim_{T \to \infty} \overline{\Phi} (T) E[U(c(T))]=0$, $\overline{\Phi} (T) \sim e^{-\kappa T}$ or faster as $T \to \infty$,
\item $r - \mu + \delta > 0$ and $ r - \alpha + \eta\rho\sigma \neq 0$
\end{enumerate}
In this paper we restrict ourselves specifically to the cases of HARA and logarithmic utility that satisfy three first conditions by definition. We also assume by default that other parameters of the problem are chosen to satisfy all the conditions above and therefore the the existence and uniqueness of the solution are guaranteed.

\subsection{Similar problems described in the literature}\label{s-o-t-a}
There are a lot of works in the framework of optimization that work with the problems that correspond to the HJB equations that look very similar to (\ref{eq:HJB21}). Almost always the authors that work with this kind of models use certain reductions of the PDE they work with. This reductions are hardly ever explained in the literature and are usually presented to the reader without any formal derivation, yet are crucial for further analysis of the problem. Before carrying out a Lie-group analysis of the problem that we have formulated in the Section \ref{problem}, let us briefly describe the works where the authors are using lie-type reductions similar to the ones that we obtain further in this paper.

In \cite{ZaripDuf} the authors formulate the problem of optimal investment with undiversifiable income risk. This problem is also an optimization problem for a portfolio of three-assets, one of which is riskless, one is risky classical stock and the third one corresponds to the nonnegative stochastic income. The problem is regarded under an infinite time horizon, so the studied equation is two-dimensional. In \cite{ZaripDufFlem} the authors regard the same problem with infinite time horizon and a specific HARA-utility. The authors use a substitution $u(x, y) = y^{\gamma} u(x/y)$, where $u$ is a value function and $x$ and $y$ are two space variables. We can see the origin of this substitution further in (\ref{inv3H3HARA}). The same substitution without the reasoning behind it is used in \cite{Zarip1999} for the problem with finite time horizon. The substitution is reducing a three-dimensional problem to a two-dimensional one.

 The authors in \cite{tebaldi} regard a problem very similar to the one described in Section \ref{problem}. They also work with a three-asset portfolio, where one asset is illiquid an is sold in a predetermined moment of time. The main difference between our approach and \cite{tebaldi} is that we work with an exogenous randomly distributed liquidation time. And again the authors in \cite{tebaldi} use a reduction $V(l, h, t) = h^{1-\gamma} V(z,t)$ but do not comment on how did the obtain this substitution.

\section{Lie-group analyses of the problem with a general liquidation time distribution and different utility functions}

After a formal maximization of (\ref{eq:Gmax21}) and (\ref{eq:Hmax21}) for the chosen utility function the equation (\ref{eq:HJB21}) becomes a three dimensional non-linear PDE (for all the details see \cite{boyazh} and \cite{BordagYamshchikovZhelezov}). As we have already said in this paper we regard two different utility functions and now we look at the cases of HARA utility and log utility separately.

\subsection{The case of HARA utility function}\label{HARA}

It is well know that a utility function $U(c)$ where the risk tolerance $R(c)$ is defined as
$R(c)=-\frac{U'(c)}{U''(c)}$ and is a linear function of $c$,
 is called  a HARA (hyperbolic absolute risk aversion) utility function.
 In this paper we use two types of utility functions: a HARA utility function $U^{HARA}(c)$ and the log-utility function $U^{LOG}(c)=\log(c)$. Let us note here that the log-utility function is often regarded as a limit case of HARA utility function. One can indeed choose HARA utility in such a way that allows a formal transition from HARA utility to log-utility  as parameter  $\gamma$ of HARA utility goes to zero, but in general this transition does not hold for any form of HARA utility. We will demonstrate this transition on different levels and because of that further in this paper we work with HARA utility in the form

	\begin{equation}\label{harautility}
		U^{HARA}(c) = \frac{1-\gamma}{\gamma}\left( \left(\frac{c}{1-\gamma}\right)^\gamma - 1\right),
	\end{equation}
with the risk tolerance $T(c)=\frac{c}{1-\gamma}$, $ 0< \gamma<1$.
One can easily see that as $\gamma \to 0$ HARA-utility function written as (\ref{harautility}) tends the to log-utility
\begin{equation} \label{HARAlimLOG}
U^{HARA}\underset{\gamma \to 0}{\longrightarrow} U^{LOG}.
\end{equation}
The HJB equation (\ref{eq:HJB21}) where we insert the HARA utility in the form (\ref{harautility})
after formal maximization procedure (see also \cite{BordagYamshchikovZhelezov}) will take the form

	\begin{eqnarray}\label{maingeneralHARA}
		 && V_t(t, l, h) +  \frac{1}{2}\eta^2h^2V_{hh} (t, l, h) + (rl + \delta h)V_l (t, l, h) + (\mu - \delta) hV_h (t, l, h)  \nonumber \\
		 &-& \frac{(\alpha - r)^2 V_{l}^2(t,l,h)+2(\alpha - r)\eta \rho h V_{l}(t,l,h) V_{lh}(t,l,h) + \eta^2 \rho^2 \sigma^2 h^2 {V_{lh}}^2 (t,l,h)}{2 \sigma^2 V_{ll}(t,l,h)} \nonumber \\
 		&+& \frac{(1-\gamma)^2}{\gamma}\overline{\Phi}(t)^{\frac{1}{1-\gamma}}V_{l} (t,l,h)^{-\frac{\gamma}{1-\gamma}}-\frac{1-\gamma}{\gamma}\overline{\Phi}(t)=0, ~~~~~ V \underset{t \to \infty}{\longrightarrow}  0.
	\end{eqnarray}
Here the investment $\pi(t, l ,h)$ and consumption $c(t, l, h)$ look as follows in terms of the value function V
\begin{eqnarray}
		 \pi (t, l ,h) &=& - \frac{\eta \rho \sigma h V_{lh}(t, l, h) + (\alpha - r) V_l (t, l, h)}{\sigma^2 V_{ll}(t,l,h)}, \label{pi:maingeneralHARA} \\
		 c(t, l, h) &=&  (1-\gamma) V_l (t, l,h)^{-\frac{1}{1-\gamma}} \overline{\Phi}(t)^{\frac{1}{1-\gamma}}. \label{c:maingeneralHARA}
\end{eqnarray}
Equation (\ref{maingeneralHARA}) is a nonlinear three dimensional PDE with three independent variables $t, h, l$. To reduce the dimension of the equation (\ref{maingeneralHARA}) we use Lie group analysis, that allows us to find the generators of the corresponding symmetry algebra admitted by this equation. In detail one can find the description of this method applied to similar PDEs in \cite{Bordag2015}.
Here we formulate the main theorem of Lie group analysis for the optimization problem with HARA type utility function.

\begin{theorem}\label{MT}
	The equation (\ref{maingeneralHARA}) admits the three dimensional Lie algebra $L^{HARA}_3$ spanned by generators $L^{HARA}_3=<\mathbf U_1, \mathbf U_2, \mathbf U_3>$, where
	
	\begin{eqnarray}
		&& \mathbf U_1 = \frac{\partial}{\partial V}, ~~~~~~\mathbf U_2 = e^{rt} \frac{\partial}{\partial l},\nonumber\\
		&& \mathbf U_3 = l \frac{\partial}{\partial l} + h \frac{\partial}{\partial h} + \left(\gamma V - (1-\gamma) \int \overline{\Phi}(t)  dt\right) \frac{\partial}{\partial V}, \label{symalharageneral}
	\end{eqnarray}
for any liquidation time distribution. Moreover, if and only if the liquidation time distribution has the exponential form, i.e. $\overline{\Phi} (t) =d e^{-\kappa t}$, where $d,\kappa$ are constants  the studied equation 	admits a four dimensional Lie algebra $L^{HARA}_4$ with an additional generator

	\begin{equation} \label{exp4sym}
	\mathbf U_4= \frac{\partial}{\partial t} -\kappa V \frac{\partial}{\partial V},
	\end{equation}
i.e. $L^{HARA}_4=<\mathbf U_1, \mathbf U_2, \mathbf U_3, \mathbf U_4>$.\\
Except finite dimensional Lie algebras (\ref{symalharageneral}) and (\ref{exp4sym}) correspondingly equation (\ref{maingeneralHARA}) admits also an infinite dimensional algebra $L_{\infty}=<\psi(h,t) \frac{\partial}{\partial V}>$ where the function $\psi(h,t)$ is any solution of the linear PDE

	\begin{equation}
 		\psi_t( h,t) +  \frac{1}{2}\eta^2h^2 \psi_{hh} ( h,t)  + (\mu - \delta) h \psi_{h} ( h,t)=0.
	\end{equation}
The Lie algebra $L^{HARA}_3$ has the following non-zero commutator relations
\begin{equation}
\left[\mathbf U_1, \mathbf U_3\right] =  \gamma \mathbf U_1,~~~~ \left[\mathbf U_2, \mathbf U_3\right] = \mathbf U_2
\end{equation}
The Lie algebra $L^{HARA}_4$ has the following non-zero commutator relations
\begin{equation}
\left[\mathbf U_1, \mathbf U_3\right] =  \gamma \mathbf U_1, ~ \left[\mathbf U_1, \mathbf U_4\right] = -\kappa \mathbf U_1, ~ \left[\mathbf U_2, \mathbf U_3\right] = \mathbf U_2, ~ \left[\mathbf U_2, \mathbf U_4\right] = -r \mathbf U_2
\end{equation}
\end{theorem}

\begin{remark}

The found Lie algebra describes the symmetry property of the equation (\ref{maingeneralHARA}) for any function $\overline{\Phi} (t)$.  In \cite{boyazh}, \cite{BordagYamshchikovZhelezov}  we have proved the theorem for existence and uniqueness of the solution of HJB equation for a liquidation time distribution which $\overline{\Phi} (t) \sim e^{-\kappa t}$ or faster as $t \to \infty$, therefore we will regard this type of the distribution studying the analytical properties of the equation further on.

\end{remark}

\begin{prof}
As in \cite{Bordag2015} we introduce the second jet bundle $j^{(2)}$ and present the equation (\ref{maingeneralHARA}) in the form
$\Delta(l,h,t,V,V_l,V_h,V_t,V_{ll},V_{ll},V_{lh},V_{hh})=0$ as a function of these variables in the jet bundle $j^{(2)}$. We look for generators of the admitted Lie algebra in the form
	
	\begin{equation}\label{operatorU}
		\mathbf U=\xi_1(l,h,t,V)\frac{\partial}{\partial l}+\xi_2(l,h,t,V)\frac{\partial}{\partial h} +\xi_3(l,h,t,V)\frac{\partial}{\partial t}+\eta_1(l,h,t,V)\frac{\partial}{\partial V},
	\end{equation}
where the functions $\xi_1,\xi_2,\xi_3,\eta_1$ can be found using the over determined system of determining equations
	
	\begin{equation}\label{operatorU2}
		\mathbf U^{(2)}\Delta(l,h,t,V,V_l,V_h,V_t,V_{ll},V_{ll},V_{lh},V_{hh})|_{\Delta=0}=0,
	\end{equation}
where $\mathbf U^{(2)}$ is the second prolongation of  $\mathbf U$ in $j^{(2)}$. We look at the action of $\mathbf U^{(2)}$ on $\Delta(l,h,t,V,V_l,V_h,V_t,V_{ll},V_{ll},V_{lh},V_{hh})$ on its solution subvariety $\Delta=0$ and obtain an overdetermined system of PDEs on the functions $\xi_1$, $\xi_2$, $\xi_3$ and $\eta_1$ from (\ref{operatorU}).
This system has 137 PDEs on the functions $\xi_1,\xi_2,\xi_3,\eta_1$. The most of them are trivial and lead to following conditions on the functions
	\begin{eqnarray}
		(\xi_1)_l &=& a_1(t),~(\xi_1)_h = 0,~(\xi_1)_V = 0, \nonumber \\
		(\xi_2)_l &=& 0,~(\xi_2)_V = 0 \nonumber \\
		(\xi_3)_l &=& 0,~(\xi_3)_h = 0,~(\xi_3)_V = 0,~ \nonumber \\
		(\eta_1)_l &=& 0,~(\eta_1)_V = d_1(h,t). \nonumber
	\end{eqnarray}
This basically means that the unknown functions have the following structure
	\begin{eqnarray}\label{condHARA}
		\xi_1(l, h, t, V) &=& a_1(t) l +a_2(t), ~~\xi_2 (l, h, t, V)=\xi_2(h,t),~~\xi_3 (l, h, t, V)=\xi_3(t), \nonumber \\
		\eta_1 (l, h, t, V)&=&d_1(h,t) V+ d_2(h,t).
	\end{eqnarray}
Here $a_1(t)$ and $d_1(h,t)$ are some functions which will be defined later.
To find the unknown functions $a_1(t), a_2(t), \xi_2(h,t), d_1(h,t), d_2(h,t)$ we should have a closer look on the non-trivial equations of the obtained system, that are left. After all simplifications we get the system of seven PDEs
\begin{eqnarray} \label{lieSystemHara}
2(\xi_2 - h {\xi_2}_h) + h {\xi_3}_t &=& 0,  \\
(1-\gamma) \overline{\Phi} \left( {\eta_1}_V - {\xi_3} - {\xi_3}_t \frac{\overline{\Phi}_t}{\overline{\Phi}} \right) + \gamma \textbf{L} (\eta_1)& = & 0, \label{1eqArb} \\
{\eta_1}_V - \gamma{\xi_1}_l - \frac{\overline{\Phi}_t}{\overline{\Phi}}{\xi_3} - (1-\gamma){\xi_3}_t &=&0, \label{2eqArb} \\
(\alpha - r){\xi_3}_t +2 \eta \rho h {\eta_1}_{hV} &=& 0, \nonumber \\
(\alpha - r)({\xi_2} - h{\xi_2}_h + h{\xi_3}_t) +\eta \rho \sigma^2 h^2 {\eta_1}_{hV} &=& 0, \nonumber \\
r{\xi_1} - {\xi_1}_t - {\xi_1}_l (\delta h + r l) +\delta {\xi_2} + (\delta h + r l) {\xi_3}_t &=& 0, \nonumber \\
(\mu - \delta) ({\xi_2} - h{\xi_2}_h +h{\xi_3}_t) - {\xi_2}_t -\frac{1}{2} \eta^2 h^2 {\eta_1}_{hh} &=& 0, \nonumber
\end{eqnarray}
where  $\textbf{L} = \frac{\partial}{\partial t} + \frac{1}{2} \eta^2 h^2 \frac{\partial^2}{\partial h^2} - (\delta - \mu)h \frac{\partial}{\partial h}$ and $\xi_1, \xi_2, \xi_3 = const $ and  $\eta_1$ satisfy (\ref{condHARA}). Using (\ref{condHARA}) we obtain a simplified system.\\
Solving the system for an arbitrary function $\overline{\Phi}(t)$ we obtain
\begin{eqnarray} \label{lieSystemHaraSol1}
\xi_1 &=& b_1 l + a_2 e^{rt},  \\
\xi_2 & = & b_1 h, \nonumber \\
\xi_3 &=&0, \nonumber \\
\eta_1 &=& b_1 \gamma V + d_2 - b_1 (1-\gamma) \int \overline{\Phi}(t)  dt + d_1(h,t). \nonumber
\end{eqnarray}
The equations (\ref{lieSystemHaraSol1}) contain three arbitrary constants $a_2, b_1, d_2$ and a function $d_1(h,t)$ which is an arbitrary solution of $\textbf{L} d_1(h,t)= 0$.
Formulas (\ref{lieSystemHaraSol1}) define three generators of finite dimensional Lie algebra $L^{HARA}_3$ (\ref{symalharageneral}) and an infinitely dimensional algebra $L_\infty$ as it was described in Theorem \ref{MT}.

If we assume that in the equations (\ref{1eqArb}) and (\ref{2eqArb}) the expression $\frac{ \overline{\Phi}_t}{ \overline{\Phi}} = const$, i.e. the liquidation time is exponentially distributed we additionally obtain the fourth symmetry (\ref{exp4sym}). It is a unique case when Lie algebra $L_3^{HARA}$ has any extensions.
\end{prof}

\subsection{The case of the log-utility function}

A logarithmic utility function is very close to the HARA-function, moreover it could be regarded as a limit case of HARA-utility (\ref{HARAlimLOG}).Yet certain properties of the logarithm make this particular case rather popular therefore we analyze it separately.

The whole approach is very similar to the method described in Section \ref{HARA} therefore we omit some details here. In the case of the log-utility function the HJB equation after the formal maximization procedure will take the following form

\begin{eqnarray}\label{maingeneralLOG}
		 && V_t(t, l, h) +  \frac{1}{2}\eta^2h^2V_{hh} (t, l, h) + (rl + \delta h)V_l (t, l, h) + (\mu - \delta) hV_h (t, l, h)  \nonumber \\
		 &-& \frac{(\alpha - r)^2 V_{l}^2(t,l,h)+2(\alpha - r)\eta \rho h V_{l}(t,l,h) V_{lh}(t,l,h) + \eta^2 \rho^2 \sigma^2 h^2 {V_{lh}}^2 (t,l,h)}{2 \sigma^2 V_{ll}(t,l,h)} \nonumber \\
 		&-& \overline{\Phi}(t)\left( \log V_l - \log \overline{\Phi}(t) + 1\right)=0, ~~~~~ V \underset{t \to \infty}{\longrightarrow}  0.
	\end{eqnarray}
	Here the investment $\pi(t, l ,h)$ and consumption $c(t, l, h)$ look as follows in terms of the value function V
\begin{eqnarray}
		 \pi (t, l ,h) &=& - \frac{\eta \rho \sigma h V_{lh}(t, l, h) + (\alpha - r) V_l (t, l, h)}{\sigma^2 V_{ll}(t,l,h)}, \label{pi:maingeneralLOG} \\
		 c(t, l, h) &=& \frac{ \overline{\Phi}(t)}{V_l (t, l, h)}.  \label{c:maingeneralLOG}
\end{eqnarray}
\begin{remark}
We choose the form of HARA-utility in such a way that (\ref{HARAlimLOG}) holds. Now we see that the maximization procedure that transforms HJB equation to PDE preserves this property as well. If we formally take a limit of (\ref{maingeneralHARA}) as $\gamma \to 0$ we obtain (\ref{maingeneralLOG}).
\end{remark}
As it turns out analogously to the previous chapter one can formulate the main theorem of Lie group analysis for this PDE.

\begin{theorem}\label{MTlog}
	The equation (\ref{maingeneralLOG}) admits the three dimensional Lie algebra $L^{LOG}_3$ spanned by generators $L^{LOG}_3=<\mathbf U_1, \mathbf U_2, \mathbf U_3>$, where
	
	\begin{eqnarray}
		&& \mathbf U_1 = \frac{\partial}{\partial V}, ~~~~~~\mathbf U_2 = e^{rt} \frac{\partial}{\partial l},\nonumber\\
		&& \mathbf U_3 = l \frac{\partial}{\partial l} + h \frac{\partial}{\partial h} -  \int \overline{\Phi}(t)  dt \frac{\partial}{\partial V}, \label{symalloggeneral}
	\end{eqnarray}
for any liquidation time distribution. Moreover, if and only if the liquidation time distribution has the exponential form, i.e. $\overline{\Phi} (t) =d e^{-\kappa t}$, where $d,\kappa$ are constants, the studied equation admits a four dimensional Lie algebra $L^{LOG}_4$ with an additional generator

	\begin{equation} \label{exp4symlog}
		\mathbf U_4= \frac{\partial}{\partial t} -\kappa V \frac{\partial}{\partial V},
	\end{equation}
i.e. $L^{LOG}_4=<\mathbf U_1, \mathbf U_2, \mathbf U_3, \mathbf U_4>$.\\
Except finite dimensional Lie algebras $L^{LOG}_3$ and $L^{LOG}_4$ correspondingly the equation (\ref{maingeneralLOG}) admits also an infinite dimensional algebra $L_{\infty}=<\psi(h,t) \frac{\partial}{\partial V}>$ where the function $\psi(h,t)$ is any solution of the linear PDE

	\begin{equation}
 		\psi_t( h,t) +  \frac{1}{2}\eta^2h^2 \psi_{hh} ( h,t)  + (\mu - \delta) h \psi_{h} ( h,t)=0.
	\end{equation}
The Lie algebra $L^{LOG}_3$ has one non-zero commutator relation $\left[\mathbf U_2, \mathbf U_3\right] =   \mathbf U_2$.\\
The Lie algebra $L^{LOG}_4$ has the following non-zero commutator relations
$$\left[\mathbf U_1, \mathbf U_4\right] = -\kappa \mathbf U_1,~~~ \left[\mathbf U_2, \mathbf U_3\right] = \mathbf U_2, ~~~\left[\mathbf U_2, \mathbf U_4\right] = -r \mathbf U_2.$$

\end{theorem}

\begin{remark}
If we compare the form of Lie algebras generators in the cases of HARA and log utilities, i.e. formulas (\ref{symalharageneral}) and (\ref{symalloggeneral}) as well as (\ref{exp4sym}) and (\ref{exp4symlog}), we can see that the formal limit procedure holds for them as well and the generators for HARA-utility transfer to generators for log-utility under a formal limit $\gamma \to 0$.
\end{remark}

\begin{prof}

Acting analogously to the proof of the Theorem \ref{MTlog} we look for the generators of the admitted Lie algebra in the form (\ref{operatorU}). A corresponding determining system obtained analogously to (\ref{operatorU2}) has 130 equations on the functions $\xi_1, \xi_2, \xi_3$ and $\eta_1$. The most of these equation are trivial and we can easily solve them. This way we obtain
	\begin{eqnarray}\label{conditionsLOG}
		(\xi_1)_l &=& b_1,~(\xi_1)_h = 0,~(\xi_1)_V = 0,  \\
		(\xi_2)_l &=& 0,~(\xi_2)_V = 0 \nonumber \\
		(\xi_3)_l &=& 0,~(\xi_3)_h = 0,~(\xi_3)_V = 0,~ \nonumber \\
		(\eta_1)_l &=& 0,~(\eta_1)_V = d_1(h,t), \nonumber
	\end{eqnarray}
 where $b_1$ is a constant and $d_1(h,t)$ is a function to be determined.
 The remaining equations can be rewritten as
\begin{eqnarray} \label{lieSystemLog}
r  {\xi_1} - (rl + \delta h) {\xi_1}_h - {\xi_1}_t +\delta {\xi_2} + (r l + \delta h){\xi_3}_t&=& 0,  \\
(\mu - \delta) ({\xi_2} - h {\xi_2}_h + h {\xi_3}_t) - {\xi_2}_t - \frac{1}{2} \eta^2 h^2 {\xi_2}_{hh} + \eta^2 h^2 {\eta_1}_{hV}& = & 0, \nonumber \\
 {\xi_2} - h {\xi_2}_h + \frac{1}{2}h {\xi_3}_t &=& 0, \nonumber \\
 {\xi_3}_t + \frac{\overline{\Phi}_t}{\overline{\Phi}} {\xi_3}  - {\eta_1}_V  &=&0, \nonumber \\
\overline{\Phi} {\xi_1}_l  \overline{\Phi}_t \log \overline{\Phi} + \overline{\Phi}(\log \overline{\Phi} - 1) {\xi_3}_t -  \overline{\Phi} \log \overline{\Phi}  {\eta_1}_V  +  \textbf{L} (\eta_1) &=& 0, \nonumber \\
 (\alpha - r) {\xi_3}_t + 2\eta \rho h  {\eta_1}_{hV} &=& 0, \nonumber \\
 (\alpha - r)( {\xi_2} - h  {\xi_2}_h + h  {\xi_3}_t) + \eta \rho \sigma^2 h^2  {\eta_1}_{hV}&=& 0, \nonumber
\end{eqnarray}
where  $\textbf{L} = \frac{\partial}{\partial t} + \frac{1}{2} \eta^2 h^2 \frac{\partial^2}{\partial h^2} - (\delta - \mu)h \frac{\partial}{\partial h}$ and $\xi_1 = \xi_1(l,t)$, $\xi_2=\xi_2(h,t)$, $\xi_3 = \xi_3 (t) $ and  $\eta_1 =\eta_1(h, t, V)$ are described in (\ref{conditionsLOG}). Inserting these functions $\xi_1, \xi_2, \xi_3$ and $\eta_1$ into (\ref{lieSystemLog}) we obtain a simplified system of determining equations.

Solving the system for an arbitrary $\overline{\Phi}(t)$ we obtain the following solution
\begin{eqnarray} \label{lieSystemLogSol}
\xi_1 &=& b_1 l + a_2 e^{rt},  \\
\xi_2 & = & b_1 h, \nonumber \\
\xi_3 &=&0, \nonumber \\
\eta_1 &=& - b_1  \int \overline{\Phi}(t)  dt + d_2 + d_1(h,t), \nonumber \\
\end{eqnarray}
where $d_1(h,t)$ is an arbitrary solution of $\textbf{L} d_1(h,t) = 0$ and $a_2, b_1$ and $d_2$ are arbitrary constants. This solution defines three different operators, that are listed in (\ref{symalloggeneral}).

If and only if the expression $\frac{\overline{\Phi}'(t)}{\overline{\Phi}(t)}$ is a constant further denoted as $\kappa$ the solution of the overdetermined system (\ref{lieSystemLog}) is as follows
\begin{eqnarray} \label{lieSystemLogExpSol}
\xi_1 &=& b_1 l + a_2 e^{rt},  \\
\xi_2 & = & b_1 h, \nonumber \\
\xi_3 &=&c_1, \nonumber \\
\eta_1 &=& - b_1  \int \overline{\Phi}(t)  dt + d_2 + d_1(h,t) - c_1 \kappa V, \nonumber
\end{eqnarray}
It means that just for the special exponential form of $\overline{\Phi}(t)$ we obtain an extension of the Lie algebra $L^{LOG}_3$ to $L^{LOG}_4$ with an additional generator $ \mathbf U_4$ (\ref{exp4symlog}).
In this way we proved the Theorem \ref{MTlog} and found the generators of $L^{LOG}_3$ and $L^{LOG}_4$ as given in (\ref{symalloggeneral}) and, correspondingly in (\ref{exp4symlog}).
\end{prof}

\section{Reductions in the general case with HARA-utility function}

In this chapter we discuss the complete set of possible reductions of three dimensional PDEs (\ref{maingeneralHARA}) arising in the case of HARA-utility function. In the previous chapter we have seen that some of the generators of the admitted Lie algebras $L^{HARA}_{3,4}$ described in Theorem \ref{MT} and $L^{LOG}_{3,4}$ as presented in Theorem \ref{MTlog} coincide in all studied cases. Let us now briefly discuss the mathematical and economic meaning of these generators.

The first generator $\mathbf U_1 = \frac{\partial}{\partial V}$ means that the original value function $V(l,h,t)$, solution of (\ref{maingeneralHARA}) for HARA utility or (\ref{maingeneralLOG}) for log utility correspondingly, can be shifted on any constant and still be a solution of the main equation. Neither allocation $\pi$ or consumption function $c$ will change their values, because the also depend only on the derivatives of the value functions (\ref{pi:maingeneralHARA}), (\ref{c:maingeneralHARA}). In some sense it is a trivial symmetry, since the equation (\ref{maingeneralHARA}) contains just the derivatives of $V(l,h,t)$ we certainly can add a constant to this function and it sill will be a solution of the equation. This symmetry does not give a rise to any reductions of the studied three dimensional PDEs.

The second generator $\mathbf U_2 = e^{rt}\frac{\partial}{\partial l}$ means that the value of the independent variable $l$ can be shifted on the arbitrary value $a e^{rt}$, i.e. the shift $l \to l+a e^{rt}$, $a-const.$ leaves the solution unaltered. From economical point of view it means that we can arbitrary shift the initial liquidity on a bank account $a, a>0$ or credit $a, a<0$. The value function $V(l,h,t)$ and the allocation-consumption strategy $(\pi,c)$ will be unaltered, see (\ref{maingeneralHARA}) or (\ref{maingeneralLOG}). This symmetry is also trivial since it does not give any reductions of the original three dimensional PDEs.

Furthermore we also get an infinitely-dimensional algebra $L_{\infty}=<\psi(h,t) \frac{\partial}{\partial V}>$ where the function $\psi(h,t)$ is any solution of the linear PDE
$\psi_t( h,t) +  \frac{1}{2}\eta^2h^2 \psi_{hh} ( h,t)  + (\mu - \delta) h \psi_{h} ( h,t)=0 $ has a very interesting meaning.
We can add any solution $\psi(h,t)$ of this equation to the value function $V(l,h,t)$ without any changes of the allocation-consumption strategy $(\pi,c)$. In economical sense it means that the additional use of some financial instrument which is the solution of  $ \psi_t( h,t) +  \frac{1}{2}\eta^2h^2 \psi_{hh} ( h,t)  + (\mu - \delta) h \psi_{h} ( h,t)=0
$, i.e. a financial instrument which value is defined just by the paper value of the illiquid asset and time, can not change the allocation-consumption strategy $(\pi,c)$.\\

Now we are going to discuss the possible reductions of the three dimensional PDE (\ref{maingeneralHARA}) arising in the case of the HARA utility in detail.

\subsection{Reductions in the case of general liquidation time distribution and HARA utility} \label{secgenhara}

In order to describe all non-equivalent invariant solutions to (\ref{maingeneralHARA}) we need to find an optimal system of subalgebras for the admitted Lie algebra (\ref{symalharageneral}) described by Theorem \ref{MT}. Let us at first remind you how does our problem look in a general case

\begin{eqnarray}\label{eq:HARA}
		 && V_t(t, l, h) +  \frac{1}{2}\eta^2h^2V_{hh} (t, l, h) + (rl + \delta h)V_l (t, l, h) + (\mu - \delta) hV_h (t, l, h)  \nonumber \\
		 &-& \frac{(\alpha - r)^2 V_{l}^2(t,l,h)+2(\alpha - r)\eta \rho h V_{l}(t,l,h) V_{lh}(t,l,h) + \eta^2 \rho^2 \sigma^2 h^2 {V_{lh}}^2 (t,l,h)}{2 \sigma^2 V_{ll}(t,l,h)} \nonumber \\
 		&+& \frac{(1-\gamma)^2}{\gamma}\overline{\Phi}(t)^{\frac{1}{1-\gamma}}V_{l} (t,l,h)^{-\frac{\gamma}{1-\gamma}}-\frac{1-\gamma}{\gamma}\overline{\Phi}(t)=0, ~~~~~ V \underset{t \to \infty}{\longrightarrow}  0.
	\end{eqnarray}
Here the investment $\pi(t, l ,h)$ and consumption $c(t, l, h)$ look as in (\ref{pi:maingeneralHARA}) and (\ref{c:maingeneralHARA}).
The Lie algebra admitted by this PDE is described in Theorem \ref{MT} and is three- or four- dimensional depending on the properties of the function $\overline{\Phi}(t)$.
The classification of all real three- and four- dimensional solvable Lie algebras is given in \cite{patera&wintern}. The authors provide optimal systems of subalgebras for every real solvable three- and four-dimensional Lie-algebra. In this Section we study a three-dimensional case. In order to have a consistency with this classification we change a basis of the corresponding algebra to a suitable one for every Lie algebras described above. One can also use the software package {\bf SymboLie} \cite{Oliveri}  (a supplement package for {\bf Mathematica}) to find an optimal system of subalgebras for a given Lie algebra directly, but we prefer to use the classification and notation provided in \cite{patera&wintern} for the sake of consistency.

The reductions can be obtained if we replace original variables with new independent and dependent variables which are invariant under the action of the Lie group or subgroup of this Lie group admitted by the equation. In this section we will list all non-equivalent reductions and provide all possible reduced equations.

First of all we should introduce the notations that we use further.

We denote by $h_i$  the subalgebras of the Lie algebra $L^{HARA}_3$ (\ref{symalharageneral}) (or $L^{HARA}_4$ correspondingly) and $H_i$ for the subgroups of the group $G^{HARA}_3$ (or $G^{HARA}_4$) which are generated with the help of the exponential map by $h_i$.

\subsubsection{System of optimal subalgebras}

At firs let us reassign the basis of $L^{HARA}_3$ to adopt the real three-dimensional Lie algebra $L^{HARA}_3$ to the classification obtained in \cite{patera&wintern} in the following way ${\mathbf U}_1 = e_2, {\mathbf U}_2 = e_1, {\mathbf U}_3 = e_3$. The basis is now defined as
\begin{equation} \label{basisL3HARA}
	e_1 = e^{rt} \frac{\partial}{\partial l}, ~ e_2 = \frac{\partial}{\partial V}, ~ e_3 = l \frac{\partial}{\partial l} + h \frac{\partial}{\partial h} + \left(\gamma V - (1-\gamma) \int \overline{\Phi}(t)  dt\right) \frac{\partial}{\partial V}.
\end{equation}

This basis has two non zero commutation relations which are given now in the following form

\begin{equation} \label{comL3HARA}
	[e_1,e_3] = e_1,~~~  [e_2, e_3]=\gamma e_2.
\end{equation}

 Now we can see that $L^{HARA}_3 = \left< e_1, e_2, e_3 \right>$ corresponds to $A^{\gamma}_{3,5}$, where $0 < \gamma < 1$, in the classification of \cite{patera&wintern}. The system of optimal subalgebras for this algebra is listed in Table \ref{TableOptSystL3HARA}.

\begin{table}[h]
\begin{center}
\begin{tabular}{|l|l|}
         \hline
         Dimension of & $~$System of optimal subalgebras of the Lie algebra $L^{HARA}_3$ (\ref{symalharageneral}) \\
         the subalgebra & \\
         \hline

         1                           & $h_1=\left< e_1 \right>, ~ h_2= \left< e_2 \right>, ~h_3=\left< e_3 \right>, ~h_4=\left< ~e_1\pm e_2 \right>$\\
  \hline
         2                           & $h_5=\left < e_1, e_2 \right>, ~h_6=\left <e_3, e_1\right>,~ h_7=\left <e_3,e_2\right>$\\

         \hline
\end{tabular}
\caption{ The optimal system of one- and two-dimensional
subalgebras of $L^{HARA}_3$ presented in (\ref{symalharageneral}). \label{TableOptSystL3HARA}}
\end{center}
\end{table}
The optimal system of one- and two- parameter subalgebras give rise to the system of one or two dimensional symmetry subgroups $H_i$ of the studied PDE. Our goal now is to find all possible corresponding reductions and to describe the solutions which are invariant under the action of the group $H_i$.

\subsubsection{One-dimensional subalgebras of $L^{HARA}_3$ and corresponding reductions} \label{sec:LHARA3}

Let us look closer at all one dimensional subalgebras listed in Table \ref{TableOptSystL3HARA}. If we try to reduce the tree dimensional PDE to a two dimensional one, then such reduction can be provided by one of the corresponding one dimensional subgroups if at all.

$L^{HARA}_3$ has four one-dimensional subalgebras $h_i$ which give rise to four one parameter subgroups $H_i$. Our goal is to study the corresponding invariant solutions. Not every subgroup out of all listed in Table \ref{TableOptSystL3HARA} provides a nontrivial reduction of the original PDE, but if a non-trivial reduction of Lie-type exists, then it can be found out through a suitable subalgebra listed in Table \ref{TableOptSystL3HARA}. We have already discussed the meaning of $h_1$ and $h_2$ above in Section \ref{secgenhara}. These two cases do not give us any reductions, as we have already discussed these two cases in the beginning go this section. Let us start with the next case.

\textbf{Case} $H_3(h_3)$. Under $h_3$ in Table \ref{TableOptSystL3HARA} we denote the subalgebra that is spanned by the generator
\begin{equation}
h_3 = < e_3 > = \left< l \frac{\partial}{\partial l} + h \frac{\partial}{\partial h} + \left(\gamma V - (1-\gamma) \int \overline{\Phi}(t)  dt\right) \frac{\partial}{\partial V} \right>. \nonumber
\end{equation}
$H_3$ denotes a corresponding subgroup. To find the invariants of $H_3$ we solve a characteristic system of the equations

\begin{equation}
\frac{dt}{0} = \frac{dl}{l} = \frac{dh}{h} = \frac{dV}{\left(\gamma V - (1-\gamma) \int \overline{\Phi}(t)  dt\right)}, \nonumber
\end{equation}
where the first equation of the system is a formal notation that shows that independent variable $t$ is actually an invariant of the equation under the action of $H_3$. We can obtain two other independent invariants solving the system above
\begin{eqnarray} \label{inv1H3HARA}
inv_1 &=& t, ~~~ inv_2 = z = \frac{l}{h},  \\
inv_3 &=& W(z,t) = h^{-\gamma} V(l,h,t) - \frac{1-\gamma}{\gamma} h^{-\gamma} \int \overline{\Phi}(t)  dt. \label{inv3H3HARA}
\end{eqnarray}
These invariants (\ref{inv1H3HARA}) can be used as new independent variables $t, z$ and the invariant (\ref{inv3H3HARA}) as the dependent variable $W(t,z)$ to reduce the three dimensional PDE (\ref{eq:HARA}) to a two dimensional one
\begin{eqnarray}\label{2dimHARAgen}
&& W_t(t, z) +  \frac{1}{2}\eta^2 \left(\gamma (\gamma - 1) W(t,z) - 2(\gamma-1) z W_{z}(t,z) + z^2 W_{zz}(t,z) \right) \nonumber \\
&+& (r z + \delta)W_z (t, z) + (\mu - \delta) (W(t,z) - z W_{z} (t,z))\nonumber \\
&-& \frac{(\alpha - r)^2 W_{z}^2(t, z) - 2(\alpha - r)\eta \rho W_{z}(t,z)((1-\gamma) W_{z}(t,z) + z W_{zz} (t,z)) }{2 \sigma^2 W_{zz}(t, z)} \nonumber \\
&-& \frac{\eta^2 \rho^2 \sigma^2 ((1-\gamma) W_{z}(t,z) + z W_{zz} (t,z))^2}{2 \sigma^2 W_{zz}(t, z)} \nonumber \\
&+& \frac{(1-\gamma)^2}{\gamma}\overline{\Phi}(t)^{\frac{1}{1-\gamma}}W_{z}^{-\frac{\gamma}{1-\gamma}} (t,z)=0, ~~~~~ W \underset{t \to \infty}{\longrightarrow}  0.
\end{eqnarray}
After this reduction the allocation $\pi(t, z ,h)$ and consumption $c(t, z, h)$ strategies (\ref{pi:maingeneralHARA}) and (\ref{c:maingeneralHARA}) look as follows
\begin{eqnarray}
		 \pi (t, z ,h) &=&h \left(\frac{\eta \rho}{\sigma}  z     +    \frac{\eta \rho \sigma (1-\gamma)  - \alpha + r}{\sigma^2} \frac{W_z (t, z)}{W_{zz}(t, z)} \right), \label{pi:HARAH3} \\
		 c(t, z, h) &=&  h (1-\gamma) W_{z}^{-\frac{1}{1-\gamma}}(t, z) \overline{\Phi}(t)^{\frac{1}{1-\gamma}}.  \label{c:HARAH3}
\end{eqnarray}

\textbf{Case} $H_4(h_4)$. This subalgebra $H_4$ is spanned by the generator $e^{rt} \frac{\partial}{\partial l} \pm\frac{\partial}{\partial V}$. We can write a characteristic system for this case
\begin{equation}\label{charsystemH4}
\frac{d l}{e^{rt}}=\frac{dh}{0}=\frac{dt}{0}=\frac{dV}{\mp1}.
\end{equation}
We can see from this system that two independent variables $t$ and $h$ are invariants. The third invariant is $W(t,h) = V(l, h, t) \mp e^{-rt} l$. This means that in this case the value function has the form $V= W(t,h) \pm e^{-rt}l $. This essentially means that $V$ in this case is a linear function of $l$. From \cite{BordagYamshchikovZhelezov} and \cite{boyazh} we know that the value function described by (\ref{valueFun}) should be concave in $l$, so this case is not interesting for our problem from the economical point of view. Since though $H_4$ gives us a reduction of the equation (\ref{maingeneralHARA}) the corresponding value function does not satisfy the conditions sufficient for a solution of (\ref{valueFun}) in the class of $l$-concave functions.

The two-dimensional subalgebras of $L^{HARA}_3$ do not give us any meaningful substitutions, what would be able to reduce the problem to an ODE.
This means that if we deal with a HARA utility function and a general form of the liquidation time distribution we have just one possibility to reduce the tree dimensional PDE (\ref{eq:HARA}) to a two dimensional one (\ref{2dimHARAgen}) using the substitution (\ref{inv1H3HARA})-(\ref{inv3H3HARA}). Any further reductions in the framework of Lie group analysis are not possible. It does not mean that any other simplifications of the PDE are not possible, but we do not have an algorithmic way to obtain it. At the same time it is important to note that if we study such problem we can for sure apply numeric or quantitate methods to study the equation (\ref{2dimHARAgen}).

\subsection{A special case of an exponential liquidation time distribution in (\ref{maingeneralHARA})}
The exponential liquidation time distribution with the survival function $\overline{\Phi}(t) = e^{-\kappa t}$ is a special case of the problem that deserves our separate attention. We have proven in Theorem \ref{MT} that in this case and just in this case we obtain an extended four dimensional Lie algebra and can hope to obtain deeper reductions than in the general case. Inserting this special form of $\overline{\Phi}(t) $ into (\ref{eq:HARA}) we obtain the following equation

	\begin{eqnarray}\label{eq:HARAexp}
		 && V_t(t, l, h) +  \frac{1}{2}\eta^2h^2V_{hh} (t, l, h) + (rl + \delta h)V_l (t, l, h) + (\mu - \delta) hV_h (t, l, h)  \nonumber \\
		 &-& \frac{(\alpha - r)^2 V_{l}^2(t,l,h)+2(\alpha - r)\eta \rho h V_{l}(t,l,h) V_{lh}(t,l,h) + \eta^2 \rho^2 \sigma^2 h^2 {V_{lh}}^2 (t,l,h)}{2 \sigma^2 V_{ll}(t,l,h)} \nonumber \\
 		&+& \frac{(1-\gamma)^2}{\gamma}e^{-\frac{\kappa t}{1-\gamma}}V_{l} (t,l,h)^{-\frac{\gamma}{1-\gamma}}-\frac{1-\gamma}{\gamma}e^{-\kappa t}=0, ~~~~~ V \underset{t \to \infty}{\longrightarrow}  0.
	\end{eqnarray}
This equation admits Lie algebra $L^{HARA}_4$ spanned by the generators (\ref{symalharageneral}) and (\ref{exp4sym}) as we have demonstrated in Theorem \ref{MT}.

Here the investment $\pi(t, l ,h)$ and consumption $c(t, l, h)$ look as follows in terms of the value function V
\begin{eqnarray}
		 \pi (t, l ,h) &=& - \frac{\eta \rho \sigma h V_{lh}(t, l, h) + (\alpha - r) V_l (t, l, h)}{\sigma^2 V_{ll}(t,l,h)}, \label{pi:HARAexp} \\
		 c(t, l, h) &=&  (1-\gamma) V_{l}(t,l,h)^{-\frac{1}{1-\gamma}} e^{\frac{-\kappa t}{1-\gamma}}.  \label{c:HARAexp}
\end{eqnarray}

Since the non-zero commutators of $L^{HARA}_4$ depend on the parameters $\kappa, r$ and $\gamma$ the inner structure of the Lie algebra $L^{HARA}_4$ is different for different values of these parameters. If we further use the classification of all solvable real three and four dimensional Lie algebras proposed in \cite{patera&wintern}, we can see that depending on the relations between the parameters of the equation we obtain two different algebraic structures. Using the notation of  \cite{patera&wintern} we see that when $\kappa \neq r \gamma$ then  $L^{HARA}_4$ corresponds to $2A_{2}$, whereas when $\kappa = r \gamma$ then  $L^{HARA}_4$ corresponds to $A^{\gamma}_{3,5} \bigoplus A_1$, see \cite{patera&wintern}. We now look at each of these cases separately.

\subsubsection{System of optimal subalgebras for $L^{HARA}_4$ for the case $\kappa \neq r \gamma$}

Let us at first regard a situation, when $\kappa \neq r \gamma$. To make the structure of $L^{HARA}_4$ visible and comparable with the notation in \cite{patera&wintern} we transform the basis of $L^{HARA}_4$ as follows
\begin{eqnarray}
e_1 &=& \frac{r \mathbf U_3 + \mathbf U_4}{\kappa - r \gamma}, ~~~ e_2 = \mathbf U_1  \nonumber\\
e_3 &=& - \frac{\kappa \mathbf U_3 + \gamma \mathbf U_4}{\kappa - r \gamma}, ~~~ e_4 = \mathbf U_2. \nonumber
\end{eqnarray}

Now the generators of $L^{HARA}_4 = \left< e_1, e_2, e_3, e_4 \right>$ have a form

\begin{eqnarray}
		&& e_1 =  \frac{r}{\kappa - r\gamma} l\frac{\partial}{\partial l} +  \frac{r}{\kappa - r\gamma} h\frac{\partial}{\partial h} + \frac{1}{\kappa - r\gamma} \frac{\partial}{\partial t} - \left( V - \frac{(1-\gamma)r}{\kappa(\kappa - r\gamma)} e^{-\kappa t} \right) \frac{\partial}{\partial V}, \nonumber\\
		&& e_2 = \frac{\partial}{\partial V}\nonumber, \\
		&& e_3 = - \frac{\kappa }{\kappa - r\gamma}l \frac{\partial}{\partial l} -  \frac{\kappa}{\kappa - r\gamma} h\frac{\partial}{\partial h} - \frac{\gamma}{\kappa - r\gamma} \frac{\partial}{\partial t} - \frac{(1-\gamma)}{\kappa - r\gamma} e^{-\kappa t} \frac{\partial}{\partial V}, \nonumber\\
		&& e_4= e^{rt} \frac{\partial}{\partial l}, \nonumber
\end{eqnarray}
where $\kappa \neq r \gamma$. In this basis the Lie algebra $L^{HARA}_4$ has only two non-zero commutation relations
\begin{equation} \label{comL3HARA}
	[e_1,e_2] = e_2,~~~  [e_3, e_4]= e_4.
\end{equation}

We can see that $L^{HARA}_4$ corresponds to $2A_{2}$ or in another common notation $A_2 \oplus A_2$, according to the classification \cite{patera&wintern}. The system of optimal subalgebras for an algebra of this type is listed in Table \ref{TableOptSystL4HARA}.

\begin{table}[h]
\begin{center}
\begin{tabular}{|l|l|}
         \hline
         Dimension of & $~$System of optimal subalgebras of algebra $L^{HARA}_4$, $\kappa \neq r \gamma$ \\
         the subalgebra & \\
         \hline

         1                       & $h_1=\left< e_2 \right>, ~ h_2= \left< e_3 \right>, ~h_3=\left< e_4 \right>, ~h_4=\left< ~e_1 + x e_3 \right>, ~h_5=\left< ~e_1 \pm e_4 \right>,$ \\
         			& $h_6=\left< ~e_2 \pm e_4 \right>, ~h_7=\left< ~e_2 \pm e_3 \right> $\\
  \hline
         2                       & $h_ 8=\left < e_1, e_3 \right>, ~h_9=\left <e_1, e_4\right>,~ h_{10}=\left <e_2,e_3\right>,~ h_{11}=\left <e_2,e_4\right>,$\\
         			& $~ h_{12}=\left <e_1+\omega e_3,e_2\right>, ~ h_{13}=\left <e_3 + \omega e_1,e_4\right>, ~ h_{14}=\left <e_1 \pm e_4,e_2\right>,$\\
				& $~ h_{15}=\left <e_3 \pm e_2,e_4\right>, ~ h_{16}=\left <e_1 + e_3,e_2 \pm e_4\right>$\\

         \hline
          3                      & $h_{17}=\left < e_1, e_3, e_2 \right>, h_{18}=\left < e_1, e_4, e_2 \right>, h_{19}=\left < e_1, e_3, e_4 \right>, h_{20}=\left < e_2, e_3, e_4 \right>, $\\
				& $h_{21}=\left < e_1 \pm e_3, e_2, e_4 \right>, h_{22}=\left < e_1 + \omega e_3, e_2, e_4 \right>$\\
         \hline
\end{tabular}
\caption{ The optimal system of one-, two- and three-dimensional subalgebras of $L^{HARA}_4$ for $\kappa \neq r \gamma$. Here $\omega$ is a parameter, $-\infty < \omega < \infty$. \label{TableOptSystL4HARA}}
\end{center}
\end{table}

We use this optimal system of subalgebras to obtain all non equivalent reductions and list them in the next section.

\subsubsection{One-dimensional subalgebras of $L^{HARA}_4$, $\kappa \neq r \gamma$} \label{neq}

$L^{HARA}_4$ in the case $\kappa \neq r \gamma$ has ten one-dimensional subalgebras listed in Table \ref{TableOptSystL4HARA}, but by far not all of them can provide meaningful reductions of (\ref{eq:HARAexp}). Before we start a step-by-step discussion regarding each subalgebra in the optimal system of subalgebras listed in Table \ref{TableOptSystL4HARA}  we should remind the reader that we have already discussed two of them before. In Section \ref{sec:LHARA3} we have already shown that subalgebras $h_1 = \left< \frac{\partial}{\partial V} \right>$ and $h_3 = \left< e^{rt} \frac{\partial}{\partial l} \right>$ and $h_6 = \left< \frac{\partial}{\partial V} \pm e^{rt} \frac{\partial}{\partial l} \right>$  describe important invariant properties of the equation (\ref{eq:HARAexp}) but they do not provide any new meaningful reductions.

Now we are going to regard other subalgebras in detail.

\textbf{Case} $H_2(h_2)$. This subalgebra is spanned by a generator
\begin{equation}\label{H2HARA}
h_2 = < e_3>  = \left< - \frac{\kappa }{\kappa - r\gamma}l \frac{\partial}{\partial l} -  \frac{\kappa}{\kappa - r\gamma} h\frac{\partial}{\partial h} - \frac{\gamma}{\kappa - r\gamma} \frac{\partial}{\partial t} - \frac{(1-\gamma)}{\kappa - r\gamma} e^{-\kappa t} \frac{\partial}{\partial V}\right>.
\end{equation}
As in the case $L^{HARA}_3$ we can find invariants of the corresponding subgroup $H_4$ solving the characteristic system

\begin{equation}
\frac{dl}{\kappa l}  = \frac{dh}{\kappa h} = \frac{dt}{\gamma} = \frac{e^{\kappa t} dV}{1-\gamma}. \nonumber
\end{equation}
Out of the characteristic system we find the following invariants
\begin{eqnarray}\label{invH2HARAEXP}
inv_1 &=& z = \frac{l}{h}, ~~~ inv_2 = \tau =  \frac{\kappa}{\gamma}t - \log h, \\
inv_3 &=& W(z,\tau) =  V + \frac{1-\gamma}{\gamma \kappa} e^{-\kappa t}. \label{invH2HARAEXP1}
\end{eqnarray}
Substituting these new independent variables $z$ and $\tau$ and a new dependent variable $W(z, \tau)$ into (\ref{maingeneralHARA}) we obtain a two-dimensional equation
\begin{eqnarray}\label{eq:HARAH2}
&&\frac{\kappa}{\gamma} W_\tau  (z, \tau) + \frac{1}{2} \eta^2 \left( 2z W_z (z, \tau) + z^2 W_{zz} (z, \tau) \right) + (rz + \delta) W_z (z, \tau) - (\mu - \delta) zW_z (z, \tau)  \nonumber\\
&-& \frac{(\alpha - r)^2 W_z^2(z, \tau) - 2 (\alpha - r) \eta \rho W_z(z,\tau) (W_z (z, \tau) + z W_{zz} (z, \tau))}{2 \sigma^2 W_{zz} (z, \tau)} \nonumber \\
&-& \frac{\eta^2 \rho^2 \sigma^2 (W_z (z, \tau) + z W_{zz} (z, \tau))^2}{2 \sigma^2 W_{zz} (z, \tau)} + \frac{(1-\gamma)^2}{\gamma} e^{-\frac{\gamma}{1-\gamma}\tau} W_z^{-\frac{\gamma}{1-\gamma}}  = 0, ~~~~~ W \underset{\tau \to \infty}{\longrightarrow}  0.
\end{eqnarray}
Here the investment $\pi(z, \tau ,h)$ and consumption $c(z, \tau, h)$ look as follows in terms of the function W
\begin{eqnarray}
		 \pi (z, \tau, h) &=&  h \left( \frac{\eta \rho}{\sigma} z + \frac{\eta \rho \sigma - \alpha + r}{\sigma^2} \frac{W_z(z, \tau)}{W_{zz}(z, \tau)} \right), \label{pi:HARAexpH2} \\
		 c(z, \tau, h) &=& h (1-\gamma) W_z(z,\tau)^{-\frac{1}{1-\gamma}}e^{-\frac{\gamma \tau}{1-\gamma}}.  \label{c:HARAexpH2}
\end{eqnarray}

\textbf{Case} $H_4(h_4)$. This sub algebra $h_4$ is spanned by the generator $e_1 = \omega e_3$
\begin{eqnarray} \label{H_4}
 h_4 &=& < e_1 + \omega e_3 >  \\
  &=& \left< \frac{r - \omega \kappa}{\kappa - r\gamma} l\frac{\partial}{\partial l} +  \frac{r - \omega \kappa }{\kappa - r\gamma} h\frac{\partial}{\partial h} + \frac{1 - \omega \gamma}{\kappa - r\gamma} \frac{\partial}{\partial t} - \left( V - \frac{(1-\gamma)(r-\omega\kappa)}{\kappa(\kappa - r\gamma)} e^{-\kappa t} \right) \frac{\partial}{\partial V} \right> .\nonumber
 \end{eqnarray}
Since parameter $\omega$ can have any value due to the interplay of the constants we need to regard three cases separately.
First, if $\omega = r/\kappa$ this case is defined by a generator
\begin{equation}
h_4=<e_1 + \frac{r}{\kappa} e_3>=\left< \frac{1}{\kappa} \frac{\partial}{\partial t} - V \frac{\partial}{\partial V} \right>. \nonumber
\end{equation}
The invariants of the corresponding subgroup $H_4$ for this case are as follows
\begin{eqnarray}\label{invH4HARAspec}
inv_1 &=& l , ~~~ inv_2 = h ,\\
inv_3 &=& W(l,h) = V e^{\kappa t}. \label{depvar:H4HARAspec}
\end{eqnarray}
Using two invariants (\ref{invH4HARAspec}) as the new independent variables and (\ref{depvar:H4HARAspec}) as the dependent variable in (\ref{eq:HARAexp}) we obtain a two dimensional PDE
\begin{eqnarray}\label{eq:H4HARAspec}
		&-&\kappa W(l, h) +  \frac{1}{2}\eta^2h^2W_{hh} (l, h) + (rl + \delta h)W_l (l, h) + (\mu - \delta) hW_h (l, h)  \nonumber \\
		 &-& \frac{(\alpha - r)^2 W_{l}^2(l,h)+2(\alpha - r)\eta \rho h W_{l}(l,h) W_{lh}(l,h) + \eta^2 \rho^2 \sigma^2 h^2 {W_{lh}}^2 (l,h)}{2 \sigma^2 W_{ll}(l,h)} \nonumber \\
 		&+& \frac{(1-\gamma)^2}{\gamma}W_{l} (l,h)^{-\frac{\gamma}{1-\gamma}}-\frac{1-\gamma}{\gamma}=0.
	\end{eqnarray}
It also means that we have a value function $V(l, h, t) = e^{- \kappa t} W(l, h)$, where $W(l,h)$ satisfies (\ref{eq:H4HARAspec}) and the time dependence of the value function $V (l, h, t)$ is defined completely by the factor $e^{- \kappa t}$ and condition $V \underset{t \to \infty}{\longrightarrow} 0$ will be satisfied for and finite $W(l,h)$.
Here the investment $\pi(l ,h)$ and consumption $c(l, h)$ look as follows in terms of the function $W$
\begin{eqnarray}
		 \pi (l ,h) &=& - \frac{\eta \rho \sigma h W_{lh}(l, h) + (\alpha - r) W_l (l, h)}{\sigma^2 W_{ll}(t,l,h)}, \label{pi:HARAexpH4spec} \\
		 c(l, h) &=&  (1-\gamma) W_l(l,h)^{-\frac{1}{1-\gamma}}.  \label{c:HARAexpH4spec}
\end{eqnarray}
The second case can be obtained if $\omega = \frac{1}{\gamma}$. In this case the algebra $h_4$ is spanned by the generator
\begin{equation}
 h_4 = < e_1 + \frac{1}{\kappa} e_3 > = \left< -\frac{1}{\gamma} l\frac{\partial}{\partial l} -  \frac{1}{\gamma} h\frac{\partial}{\partial h} - \left( V + \frac{1-\gamma}{\kappa\gamma} e^{-\kappa t} \right) \frac{\partial}{\partial V} \right> . \nonumber
\end{equation}
One can find the following invariants of the subgroup $H_4$
\begin{eqnarray}\label{invH4HARAEXPspec}
inv_1 &=& z = \frac{l}{h}, ~~~ inv_2 = t, \\
inv_3 &=& W(z,t) =  h^{-\gamma}V + \frac{1-\gamma}{\gamma \kappa} h^{-\gamma} e^{-\kappa t}. \label{depvar:H4HARAEXPspec}
\end{eqnarray}
Substituting the invariants $z$  and $t$ (\ref{invH4HARAEXPspec}) as independent variables and the invariant $W(z,t)$ (\ref{depvar:H4HARAEXPspec}) as the dependent variable into (\ref{eq:HARAexp}) we derive the following two dimensional equation
\begin{eqnarray}\label{eq:HARAH4spec}
&& W_t(t, z) +  \frac{1}{2}\eta^2 \left(\gamma (\gamma - 1) W(t,z) - 2(\gamma-1) z W_{z}(t,z) + z^2 W_{zz}(t,z) \right) \nonumber \\
&+& (r z + \delta)W_z (t, z) + (\mu - \delta) (W(t,z) - z W_{z} (t,z))\nonumber \\
&-& \frac{(\alpha - r)^2 W_{z}^2(t, z) - 2(\alpha - r)\eta \rho W_{z}(t,z)((1-\gamma) W_{z}(t,z) + z W_{zz} (t,z)) }{2 \sigma^2 W_{zz}(t, z)} \nonumber \\
&-& \frac{\eta^2 \rho^2 \sigma^2 ((1-\gamma) W_{z}(t,z) + z W_{zz} (t,z))^2}{2 \sigma^2 W_{zz}(t, z)} \nonumber \\
&+& \frac{(1-\gamma)^2}{\gamma} e^{-\frac{\kappa t}{1-\gamma}} W_z^{-\frac{\gamma}{1-\gamma}}  = 0,  ~~~W \underset{t \to \infty}{\longrightarrow} 0.
\end{eqnarray}
Here the investment $\pi(t, z ,h)$ and consumption $c(t, z, h)$ look as follows in terms of the function $W$
\begin{eqnarray}
		 \pi (t, z ,h) &=& h \left( \frac{\eta \rho}{\sigma} z + \frac{\eta \rho \sigma (1 - \gamma)- \alpha + r}{\sigma^2} \frac{W_z (t, z)}{W_{zz}(t,z)} \right), \label{pi:HARAexpH4spec2} \\
		 c(t, z, h) &=&  h(1-\gamma)  e^{\frac{-\kappa t}{1-\gamma}} W_{z}(t,z)^{-\frac{1}{1-\gamma}}.  \label{c:HARAexpH4spec2}
\end{eqnarray}
Before we move on, we need to point out two facts about this substitution. First of all, it directly corresponds to the substitution (\ref{inv1H3HARA}) that we have found earlier for a case of general liquidation time distribution. Second notion that we need to make is that an analogous symmetry is used in \cite{tebaldi}. The framework of the problem is a bit different, since authors regard a fixed pre-determined liquidation time, yet the substitution they use to reduce a three dimensional PDE to a two dimensional one is very similar. The authors do not carry out a complete analysis of their problem and have to work with an equation of the second order. This makes a problem significantly more complicated, yet in their framework a Lie-type reduction to a one dimensional equation is possible, as it was shown in \cite{boyazh}. The authors in \cite{ZaripDufFlem} work with the problem of random income in an infinite time set-up, so their problem is two-dimensional by design. They also use a similar substitution to reduce their two dimensional problem to a one dimensional case.  Yet it is clear that the substitution they use corresponds to this one. In fact in \cite{boyazh} it is shown how the three-dimensional problem with an illiquid asset corresponds to a two dimensional problem of random income with infinite time horizon. Now, since we have carried out a complete analysis of all Lie-type substitutions we an see which of them were explored in the literature before and give a solid mathematical explanation of this substitutions instead of an educated guess that is most commonly used to simplify the problems of such type

These two cases $\omega = r/\kappa$ and $\omega = \frac{1}{\gamma}$ are special since the generators, that span $H_4$, differ significantly form the generator in the most general case (\ref{H_4}), when $\omega \neq \frac{r}{\kappa}, \frac{1}{\gamma}$.

The invariants in this more general case are as follows

\begin{eqnarray}\label{invH4HARAEXP}
inv_1 &=& z = \frac{l}{h}, ~~~ inv_2 = \tau =  \frac{r - \omega \kappa}{1 - \omega \gamma} t - \log h, \\
inv_3 &=& W(z,\tau) =  V e^{ \frac{\kappa- r \gamma}{1-\omega \gamma} t} + \frac{1-\gamma}{\gamma \kappa} e^{\frac{ \gamma(\omega \kappa - r)}{1- \omega \gamma} t}, ~~~ \omega \neq \frac{r}{\kappa}, \frac{1}{\gamma}. \label{depvar:H4HARAEXP}
\end{eqnarray}
As in previous cases we chose the first two invariants as independent variables and the last invariant $W(z, \tau)$ defined as (\ref{depvar:H4HARAEXP}) as a new dependent variable. Substituting these variables into (\ref{eq:HARAexp}) we obtain the reduced two dimensional equation
\begin{eqnarray}\label{eq:HARAH4}
&&-\frac{\kappa - r \gamma}{1 - \omega \gamma}W(z, \tau) + \frac{r-\omega \kappa}{1-\omega\gamma} W_\tau  (z, \tau) + \frac{1}{2} \eta^2 \left( 2z W_z (z, \tau) + z^2 W_{zz} (z, \tau) \right) \nonumber \\
&+& (rz + \delta) W_z (z, \tau) - (\mu - \delta) z W_z (z, \tau)  \nonumber\\
&-& \frac{(\alpha - r)^2 W_z^2(z, \tau) - 2 (\alpha - r) \eta \rho W_z (z, \tau) (W_z (z, \tau) + z W_{zz} (z, \tau))}{2 \sigma^2 W_{zz} (z, \tau)}\nonumber \\
&-&\frac{\eta^2 \rho^2 \sigma^2 (W_z (z, \tau) + z W_{zz} (z, \tau))^2}{2 \sigma^2 W_{zz} (z, \tau)} + \frac{(1-\gamma)^2}{\gamma}  W_z^{-\frac{\gamma}{1-\gamma}}  = 0.
\end{eqnarray}
Indeed, the condition on $V$ is rewritten as $W \underset{\tau \to \infty}{\longrightarrow} 0$.
Here the investment $\pi(z, \tau ,h)$ and consumption $c(z, \tau, h)$ look as follows in terms of the value function W
\begin{eqnarray}
		 \pi (z, \tau ,h) &=& \left( \frac{\eta \rho}{\sigma} z + \frac{\eta \rho \sigma - \alpha + r}{\sigma^2} \frac{W_z (z, \tau)}{ W_{zz}(z, \tau)} \right), \label{pi:HARAexpH4} \\
		 c(z, \tau, h) &=&  (1-\gamma) h  e^{-\frac{\gamma \tau}{1-\gamma}} W_{z}^{-\frac{1}{1-\gamma}}.  \label{c:HARAexpH4}
\end{eqnarray}

\textbf{Case} $H_5(h_5)$. This subalgebra is spanned by the following generator
\begin{eqnarray}
h_5 &=& < e_1 \mp e_4 > \\
&=&\left< \left( \frac{r}{\kappa - r\gamma} l \pm e^{rt} \right) \frac{\partial}{\partial l} +  \frac{r}{\kappa - r\gamma} h\frac{\partial}{\partial h} + \frac{1}{\kappa - r\gamma} \frac{\partial}{\partial t} - \left( V - \frac{(1-\gamma)r}{\kappa(\kappa - r\gamma)} e^{-\kappa t} \right) \frac{\partial}{\partial V} \right>. \nonumber
\end{eqnarray}
Solving the corresponding characteristic system we find the following invariants of the subgroup $H_5$
\begin{eqnarray}\label{invH5HARAEXP}
inv_1 &=& x = l e^{-r t} \mp (\kappa -r \gamma)t , ~~~ inv_2 = y =  h e^{-rt}, \\
inv_3 &=& W(x,y) =  V e^{(\kappa - r \gamma)t} - \frac{1-\gamma}{\gamma \kappa} e^{- r \gamma t}. \label{depvar:H5HARAEXP}
\end{eqnarray}

Using expressions (\ref{invH5HARAEXP}) as new independent variables $z$ and $\tau$ and (\ref{depvar:H5HARAEXP}) as a new dependent variable $W(x, y)$ we reduce (\ref{eq:HARAexp}) to the two dimensional PDE

 \begin{eqnarray}\label{eq:HARAH5}
		 && -(\kappa - r \gamma)W(x,y)+  \frac{1}{2}\eta^2y^2W_{y y} (x,y) + (\delta y \mp (\kappa - r\gamma))W_x (x,y) + (\mu - \delta) y W_y (x,y)  \nonumber \\
		 &-& \frac{(\alpha - r)^2 W_{x}^2(x,y)+2(\alpha - r)\eta \rho y W_{x}(x,y) W_{xy}(x,y) + \eta^2 \rho^2 \sigma^2 y^2 {W_{xy}}^2 (x,y)}{2 \sigma^2 W_{xx}(x,y)} \nonumber \\
 		&+& \frac{(1-\gamma)^2}{\gamma} W_{x}^{-\frac{\gamma}{1-\gamma}}(x,y)=0, ~~~W(x,0) \underset{x \to \mp \infty}{\longrightarrow} 0.
	\end{eqnarray}
Here the investment $\pi(x, y ,h)$ and consumption $c(x, y, h)$ look as follows in terms of the value function W
\begin{eqnarray}
		 \pi (x, y ,h) &=& - h \frac{\eta \rho \sigma W_{xy}(x, y) + (\alpha - r) y^{-1} W_x (x, y)}{\sigma^2 W_{xx}(x,y)}, \label{pi:HARAexpH5} \\
		 c(x, y, h) &=& h (1-\gamma) y^{-1} W_{x}^{-\frac{1}{1-\gamma}}.  \label{c:HARAexpH5}
\end{eqnarray}
\textbf{Case} $H_7(h_7)$. The last one dimensional subalgebra listed in Table \ref{TableOptSystL4HARA} is subalgebra $h_7$ spanned by
\begin{equation}
h_7 = < e_2 \mp e_3 > = \left< \mp \frac{\kappa}{\kappa - r\gamma} l \frac{\partial}{\partial l} \mp  \frac{\kappa}{\kappa - r\gamma} h\frac{\partial}{\partial h} \mp \frac{\gamma}{\kappa - r\gamma} \frac{\partial}{\partial t} - \left( 1 \mp \frac{1-\gamma}{\kappa - r\gamma} e^{-\kappa t} \right) \frac{\partial}{\partial V} \right> . \nonumber
\end{equation}
Solving the characteristic system we find the following invariants of the corresponding subgroup $H_7$

\begin{eqnarray}\label{invH7HARAEXP}
inv_1 &=& z = \frac{l}{h}, ~~~ inv_2 = \tau =  \frac{\kappa}{\gamma}t - \log h,  \\
inv_3 &=& W(z,\tau) =  V \pm \frac{\kappa -r\gamma}{\gamma}t + \frac{1-\gamma}{\gamma \kappa} e^{- \kappa t}.\label{depvar:H7HARAEXP}
\end{eqnarray}

As before to get solutions of (\ref{eq:HARAexp}) invariant under the action of $H_7$ we use the invariants (\ref{invH7HARAEXP}) as independent variables $z, \tau$ and the invariant (\ref{depvar:H7HARAEXP}) as the dependent variable $W(z,\tau)$. This way we reduce equation (\ref{eq:HARAexp}) to a two dimensional PDE that looks as follows

 \begin{eqnarray}\label{eq:HARAH7}
		&&\frac{\kappa}{\gamma}W_{\tau}(z, \tau) \pm \frac{\kappa - r \gamma}{\gamma}+ \frac{1}{2} \eta^2 \left( 2z W_z (z, \tau) + z^2 W_{zz} (z, \tau) \right) + (rz + \delta) W_z (z, \tau)  \\
&-& (\mu - \delta) zW_z (z, \tau) - \frac{(\alpha - r)^2 W_z^2(z, \tau) - 2 (\alpha - r) \eta \rho W_z(z,\tau) (W_z (z, \tau) + z W_{zz} (z, \tau))}{2 \sigma^2 W_{zz} (z, \tau)} \nonumber \\
&-& \frac{\eta^2 \rho^2 \sigma^2 (W_z (z, \tau) + z W_{zz} (z, \tau))^2}{2 \sigma^2 W_{zz} (z, \tau)} + \frac{(1-\gamma)^2}{\gamma} e^{-\frac{\gamma}{1-\gamma}\tau} W_z^{-\frac{\gamma}{1-\gamma}}  = 0, ~W(z,\tau) \underset{\tau \to \infty}{\longrightarrow} 0. \nonumber
	\end{eqnarray}

Here the investment $\pi(z, \tau ,h)$ and consumption $c(z, \tau , h)$ look as follows in terms of the function W
\begin{eqnarray}
		 \pi (z, \tau, h) &=&  h \left( \frac{\eta \rho}{\sigma}z + \frac{\eta \rho \sigma - \alpha + r}{\sigma^2} \frac{W_z(z, \tau)}{W_{zz}(z, \tau)} \right), \label{pi:HARAexpH7} \\
		 c(z, \tau, h) &=&  (1-\gamma) h W_z(z,\tau)^{-\frac{1}{1-\gamma}}e^{-\frac{\gamma}{1-\gamma} \tau}. \label{c:HARAexpH7}
\end{eqnarray}
We studied in detail all seven one dimensional subalgebras from optimal system of subalgebras which can describe non equivalent invariant solutions of (\ref{eq:HARA}) in the case $\kappa \neq r \gamma$. We demonstrated that only in four cases meaningful reductions to a two dimensional PDEs are possible.

\subsubsection{Two-dimensional subalgebras of $L^{HARA}_4$}

As we studied one dimensional subalgebras we obtained the invariant solutions which are unaltered under the action of a one parameter group generated by one dimensional subalgebras from the system of optimal subalgebras. Now we are going to find all non equivalent invariant solutions which are invariant under actions of two parameter subalgebras. This gives us a possibility to reduce a three dimensional PDE to an ODE.

In this section we go further and looking at the second row of the Table \ref{TableOptSystL4HARA} find the deeper reductions that can reduce PDE (\ref{eq:HARA}) to an ordinary differential equation.

\textbf{Case} $H_8(h_8)$. The first two dimensional sub algebra listed in Table \ref{TableOptSystL4HARA} is subalgebra $h_8 = < e_1, e_3 >$ spanned by two generator defined as follows
\begin{eqnarray}
e_1  &=&   \frac{r}{\kappa - r\gamma} l\frac{\partial}{\partial l} +  \frac{r}{\kappa - r\gamma} h\frac{\partial}{\partial h} + \frac{1}{\kappa - r\gamma} \frac{\partial}{\partial t} - \left( V - \frac{(1-\gamma)r}{\kappa(\kappa - r\gamma)} e^{-\kappa t} \right) \frac{\partial}{\partial V}, \nonumber \\
e_3  &=& - \frac{\kappa }{\kappa - r\gamma}l \frac{\partial}{\partial l} -  \frac{\kappa}{\kappa - r\gamma} h\frac{\partial}{\partial h} - \frac{\gamma}{\kappa - r\gamma}
\frac{\partial}{\partial t} - \frac{(1-\gamma)}{\kappa - r\gamma} e^{-\kappa t} \frac{\partial}{\partial V}, ~~~ \kappa \neq r \gamma.\nonumber
\end{eqnarray}
The two dimensional subalgebra is spanned by two generators and each of them was actually studied before. We should find a simultaneous solution to both characteristic systems. We can use our previous knowledge and reformulate one of the characteristic systems in terms of invariant variables of the other one.  This could be done in the following way. We have found a general form of the invariants of $e_1$ in case $H_2(h_2)$ (\ref{invH2HARAEXP}), (\ref{invH2HARAEXP1}).  Let us list them here again for the convenience of the reader
\begin{eqnarray}
inv_1 &=& z = \frac{l}{h}, ~~~ inv_2 = \tau =  \frac{\kappa}{\gamma}t - \log h, \nonumber \\
inv_3 &=& W(z,\tau) =  V + \frac{1-\gamma}{\gamma \kappa} e^{-\kappa t}. \nonumber
\end{eqnarray}
If we rewrite the second generator of the subalgebra $h_8$ in terms of these three invariants $z, \tau$ and $W$ as new independent and dependent variables correspondingly, we obtain
\begin{equation}
e_3 = \frac{1}{\gamma} \frac{\partial}{\partial \tau} - W \frac{\partial}{\partial W}.
\end{equation}
Solving a corresponding characteristic system $\frac{d \tau}{1/\gamma} = \frac{d W}{- W}$ we obtain a new invariant
\begin{equation}
inv_{e_3} = Y(z) =  W(z, \tau) e^{\gamma \tau}.
\end{equation}
This way we obtain an invariant solution under the action of two parameter subgroup $H_8$. It means that we can take $Y(z)$ as a new dependent variable in (\ref{eq:HARAH2}) and $z$ as a new independent one. Substituting these invariants into PDE (\ref{eq:HARAH2}) we obtain a new ODE
\begin{eqnarray}\label{eq:HARAH8}
&-&\kappa Y (z) + \frac{1}{2} \eta^2 \left( 2z Y_z (z) + z^2 Y_{zz} (z) \right) + (rz + \delta) Y_z (z) - (\mu - \delta) z Y_z (z)  \nonumber\\
&-& \frac{(\alpha - r)^2 Y_z^2(z) - 2 (\alpha - r) \eta \rho Y_z(z) (Y_z (z) + z Y_{zz} (z)) + \eta^2 \rho^2 \sigma^2 (Y_z (z) + z Y_{zz} (z))^2}{2 \sigma^2 Y_{zz} (z)} \nonumber \\
&+& \frac{(1-\gamma)^2}{\gamma} Y_z^{-\frac{\gamma}{1-\gamma}} (z) = 0.
\end{eqnarray}
The condition $W(z,\tau) \underset{\tau \to \infty}{\longrightarrow} 0$ is satisfied for each finite solution $Y(z)$, because $W(z, \tau) = e^{- \gamma \tau} Y(z) $
Naturally, the investment $\pi(z, \tau ,h)$ and consumption $c(z, \tau , h)$ in terms of $Y(z)$ look now as follows

\begin{eqnarray}
		 \pi (z, \tau ,h) &=& h \left(\frac{\eta \rho}{\sigma}z + \frac{\eta \rho \sigma  - \alpha + r}{\sigma^2}  \frac{Y_z (z) }{Y_{zz}(z) } \right), \label{pi:HARAexpH8} \\
		 c(z, \tau, h) &=& h (1-\gamma) Y_z(z)^{-\frac{1}{1-\gamma}}.  \label{c:HARAexpH8}
\end{eqnarray}
In terms of original variables $l, h, t$ and $V$ the substitution looks as follows
\begin{eqnarray}
z &=& \frac{l}{h}, ~~~ \tau =  \frac{\kappa}{\gamma}t - \log h, \label{2subHARAH8}  \\
Y(z) &=& \left( V  e^{\kappa t} + \frac{1-\gamma}{\gamma \kappa} \right) h^{- \gamma}. \nonumber
\end{eqnarray}
It also means that if we obtain a solution $Y(z)$ for (\ref{eq:HARAH8}) we obtain the value function that in terms of original variables looks like
$$
V(t, l ,h) = e^{-\kappa t} h^{\gamma} Y(l/h) - \frac{1-\gamma}{\gamma \kappa} e^{- \kappa t},
$$
and the condition $V(l, h, t) \underset{t \to \infty}{\longrightarrow} 0$ is satisfied.

All other two and three dimensional subalgebras listed in Table \ref{TableOptSystL4HARA} do not give a reduction of the original equation (\ref{eq:HARA}) to an ODE, so we will not regard them in detail.

\subsubsection{System of optimal subalgebras. Case $\kappa=r \gamma$, i.e. $L^{HARA}_4=A^{\gamma}_{3,5} \bigoplus A_1$}
In Section \ref{neq} we worked with the case $\kappa \neq r \gamma$ now we study the special case $\kappa = r \gamma$.
When $\kappa = r \gamma$ as we have mentioned above $L^{HARA}_4$ has a different structure according to the classification \cite{patera&wintern}. We substitute now $\kappa=r \gamma$ into (\ref{eq:HARAexp}) and regard the following equation

\begin{eqnarray}\label{eq:HARAexpGamma}
		 && V_t(t, l, h) +  \frac{1}{2}\eta^2h^2V_{hh} (t, l, h) + (rl + \delta h)V_l (t, l, h) + (\mu - \delta) hV_h (t, l, h)   \\
		 &-& \frac{(\alpha - r)^2 V_{l}^2(t,l,h)+2(\alpha - r)\eta \rho h V_{l}(t,l,h) V_{lh}(t,l,h) + \eta^2 \rho^2 \sigma^2 h^2 {V_{lh}}^2 (t,l,h)}{2 \sigma^2 V_{ll}(t,l,h)} \nonumber \\
 		&+& \frac{(1-\gamma)^2}{\gamma}e^{-\frac{r \gamma t}{1-\gamma}}V_{l} (t,l,h)^{-\frac{\gamma}{1-\gamma}}-\frac{1-\gamma}{\gamma}e^{-r \gamma t}=0, ~~~ V(l, h, t) \underset{t \to \infty}{\longrightarrow} 0.\nonumber
	\end{eqnarray}

Here the investment $\pi(t, l ,h)$ and consumption $c(t, l, h)$ look as follows in terms of the value function $V(t , l, h)$
\begin{eqnarray}
		 \pi (t, l ,h) &=& - \frac{\eta \rho \sigma h V_{lh}(t, l, h) + (\alpha - r) V_l (t, l, h)}{\sigma^2 V_{ll}(t,l,h)}, \label{pi:HARAexpG} \\
		 c(t, l, h) &=&  (1-\gamma) V^{-\frac{1}{1-\gamma}}_l e^{- \frac{r \gamma }{1-\gamma} t}.  \label{c:HARAexpG}
\end{eqnarray}

In this case we transform the old basis of $L^{HARA}_4=< \mathbf U_1, \mathbf U_2, \mathbf U_3, \mathbf U_4>$ described in (\ref{symalharageneral}) and (\ref{exp4sym}) into the following one
$$e_1 =  \mathbf U_2, ~ e_2 = \mathbf U_1, ~ e_3 = - \frac{1}{r} \mathbf U_4, ~ e_4 = \mathbf U_3 +  \frac{1}{r} \mathbf U_4,$$
where we use the relation $\kappa = r \gamma$.
Now the new basis looks like this

\begin{eqnarray} \label{basisWinternL4HARA}
		&& e_1 =  e^{rt} \frac{\partial}{\partial l}     \\
		&& e_2 = \frac{\partial}{\partial V}\nonumber \\
		&& e_3 = - \frac{1}{ r} \frac{\partial}{\partial t} + \gamma V \frac{\partial}{\partial V} \nonumber\\
		&& e_4= l\frac{\partial}{\partial l} + h\frac{\partial}{\partial h} + \frac{1}{r} \frac{\partial}{\partial t} + \frac{1-\gamma}{r \gamma} e^{-r \gamma t}  \frac{\partial}{\partial V} \nonumber
\end{eqnarray}
In this basis there are only two non-zero commutation relations
\begin{equation} \label{comL3HARA}
	[e_1,e_3] = e_1,~~~  [e_2, e_3]= \gamma e_2.
\end{equation}

 Now we can see that $L^{HARA}_4$ corresponds to the algebra of the type $A^{\gamma}_{3,5} \bigoplus A_1$, in the notation of \cite{patera&wintern}. The system of optimal subalgebras for this algebra is listed in Table \ref{TableOptSystL4HARA2}.

\begin{table}[h]
\begin{center}
\begin{tabular}{|l|l|}
         \hline
         Dimension of & $~~~~~~~~~~~~~~~~~~~~~$System of optimal subalgebras of algebra $L^{HARA}_4$ \\
         the subalgebra & \\
         \hline

         1                       & $h_1=\left< e_1 \right>, ~ h_2= \left< e_2 \right>, ~h_3=\left< e_4 \right>, ~h_4=\left< ~e_1 \pm e_4 \right>, ~h_5=\left< ~e_2 \pm e_4 \right>,$ \\
         			& $h_6=\left< ~e_3 + \omega e_4 \right>, ~h_7=\left< ~e_1 \pm e_2 + \omega e_4 \right> $\\
  \hline
         2                       & $h_ 8=\left < e_1, e_2 \right>, ~h_9=\left <e_1, e_4\right>,~ h_{10}=\left <e_2,e_4\right>,~ h_{11}=\left <e_3,e_4\right>,$\\
         			& $~ h_{12}=\left <e_1 \pm e_2,e_4\right>, ~ h_{13}=\left <e_1, e_2 \pm e_4\right>, ~ h_{14}=\left <e_1 \pm e_4,e_2 + \omega e_4\right>,$\\
				& $~ h_{15}=\left <e_3 + \omega e_4,e_1\right>, ~ h_{16}=\left <e_3 + \omega e_4,e_2 \right>$\\

         \hline
          3                      & $h_{17}=\left < e_1, e_2, e_4 \right>, h_{18}=\left < e_3, e_4, e_1 \right>, h_{19}=\left < e_3, e_4, e_2 \right>, h_{20}=\left < e_3 + \omega e_4, e_1, e_2 \right>$\\
         \hline
\end{tabular}
\caption{ The optimal system of one-, two- and three-dimensional
subalgebras of $L^{HARA}_4$, if $\kappa = r \gamma$,  where $\omega$ is a parameter such that $-\infty < \omega < \infty$. \label{TableOptSystL4HARA2}}
\end{center}
\end{table}

\subsubsection{One-dimensional subalgebras of $L^{HARA}_4$}

We use now the Table \ref{TableOptSystL4HARA2} to list all non equivalent possible reductions of the three dimensional PDE (\ref{eq:HARAexpGamma}) in the the case $\kappa = r \gamma$. As we have shown before the first two subgroups $H_1$ and $H_2$, spanned by $h_1 = \left< e^{rt} \frac{\partial }{\partial l}\right>$ and $h_2 = \left< \frac{\partial }{\partial W}\right>$ correspondingly, provide no meaningful reductions of (\ref{eq:HARAexp}) or, correspondingly, (\ref{eq:HARAexpGamma}). Let us move on to the next case.

\textbf{Case} $H_3(h_3)$. This subalgebra is spanned by $e_4$, i.e.
\begin{equation}
h_3 = < e_4 > = \left< l\frac{\partial}{\partial l} + h\frac{\partial}{\partial h} + \frac{1}{r} \frac{\partial}{\partial t} + \frac{1-\gamma}{r \gamma} e^{-r \gamma t}  \frac{\partial}{\partial V} \right>. \nonumber
\end{equation}
Solving the corresponding characteristic system we find the following invariants of the subgroup $H_3$
\begin{eqnarray}\label{invH3HARAspecEXP}
inv_1 &=& z = \frac{l}{h}, ~~~ inv_2 = \tau =  r t - \log h, \\
inv_3 &=& W(z,\tau) =  V + \frac{1 -\gamma}{r\gamma^2} e^{- r \gamma t}.\label{depvar:H3HARAspecEXP}
\end{eqnarray}
Substituting these invariants into (\ref{eq:HARAexpGamma}) we obtain a two dimensional PDE which describes all solutions that are invariant under the action of the subgroup $H_3$
 \begin{eqnarray}\label{eq:HARAspecH3}
&&r W_\tau  (z, \tau) + \frac{1}{2} \eta^2 \left( 2z W_z (z, \tau) + z^2 W_{zz} (z, \tau) \right) + (rz + \delta) W_z (z, \tau) - (\mu - \delta) zW_z (z, \tau)  \nonumber\\
&-& \frac{(\alpha - r)^2 W_z^2(z, \tau) - 2 (\alpha - r) \eta \rho W_z(z,\tau) (W_z (z, \tau) + z W_{zz} (z, \tau))}{2 \sigma^2 W_{zz} (z, \tau)}  \\
&-& \frac{\eta^2 \rho^2 \sigma^2 (W_z (z, \tau) + z W_{zz} (z, \tau))^2}{2 \sigma^2 W_{zz} (z, \tau)} + \frac{(1-\gamma)^2}{\gamma} e^{-\frac{\gamma}{1-\gamma}\tau} W_z^{-\frac{\gamma}{1-\gamma}}  = 0, ~~~ W(z, \tau) \underset{\tau \to \infty}{\longrightarrow} 0. \nonumber
\end{eqnarray}
Here the investment $\pi(z, \tau ,h)$ and consumption $c(z, \tau, h)$ look as follows in terms of the value function W
\begin{eqnarray}
  \label{pi:HARAexpH2} \\
		 \pi (z, \tau, h) &=&  h \left(\frac{\eta \rho}{\sigma}z+ \frac{\eta \rho \sigma - \alpha + r}{\sigma^2} \frac{W_z(z, \tau)}{W_{zz}(z, \tau)} \right), \label{pi:HARAexpGH3} \\
		 c(z, \tau, h) &=& h (1-\gamma) W_z(z,\tau)^{-\frac{1}{1-\gamma}}e^{-\frac{\gamma}{1-\gamma} \tau}.  \label{c:HARAexpGH3}
\end{eqnarray}

\textbf{Case} $H_4(h_4)$. As one can see in the Table \ref{TableOptSystL4HARA2} this subalgebra is spanned by
\begin{equation}
h_4 = < e_1 \pm e_4 > = \left< (l \pm e^{rt})\frac{\partial}{\partial l} + h\frac{\partial}{\partial h} + \frac{1}{r} \frac{\partial}{\partial t} + \frac{1-\gamma}{r \gamma} e^{-r \gamma t}  \frac{\partial}{\partial V} \right>. \nonumber
\end{equation}
We find the following invariants solving the characteristic system of equations

\begin{eqnarray}\label{invH4HARAspecEXP}
inv_1 &=& x = l e^{-rt} \mp rt, ~~~ inv_2 = y =  h e^{-rt}, \\
inv_3 &=& W(x, y) =  V + \frac{1 -\gamma}{r\gamma^2} e^{- r \gamma t}. \label{depvar:H4HARAspecEXP}
\end{eqnarray}

We use the invariants $z$ and $\tau$ (\ref{invH4HARAspecEXP}) of $H_4$ as independent variables and the invariant $W(z, \tau)$ (\ref{depvar:H4HARAspecEXP}) as the dependent variable. This way we reduce equation (\ref{eq:HARAexpGamma}) to a two dimensional PDE that looks as follows

 \begin{eqnarray}\label{eq:HARAspecH4}
&\mp& r  W_x (x, y)+ \delta y W_x (x, y)+ (\mu -\delta)y W_{y} (x, y) + \frac{1}{2}\eta^2 y^2 W_{yy}(x, y)   \nonumber\\
&-& \frac{(\alpha - r)^2 W_x^2(x, y) - 2(\alpha - r) \eta \rho y W_x (x, y)W_{x y}(x, y) + \eta^2 \rho^2 \sigma^2 y^2 W_{x y}^2(x, y)}{2\sigma^2W_{xx}(x, y)} \nonumber\\
&+& \frac{(1-\gamma)^2}{\gamma} W_x^{-\frac{\gamma}{1-\gamma}}  (x, y) = 0, ~~~W(x,0) \underset{x \to \mp \infty}{\longrightarrow} 0.
\end{eqnarray}
Here the investment $\pi(x, y ,h)$ and consumption $c(x, y, h)$ look as follows in terms of the value function W
\begin{eqnarray}
		 \pi (x, y ,h) &=& - h \frac{\eta \rho \sigma W_{xy}(x, y) + (\alpha - r) y^{-1} W_x (x, y)}{\sigma^2 W_{xx}(x, y)}, \label{pi:HARAexpGH4} \\
		 c(x, y, h) &=&  h (1-\gamma) y^{-1} W_x^{-\frac{1}{1-\gamma}}(x,y).  \label{c:HARAexpGH4}
\end{eqnarray}

\textbf{Case} $H_5(h_5)$. This subalgebra is spanned by
\begin{equation}
h_5 = < e_2 \pm e_4 > = \left< l \frac{\partial}{\partial l} + h\frac{\partial}{\partial h} + \frac{1}{r} \frac{\partial}{\partial t} + \left(\pm 1 + \frac{1-\gamma}{r \gamma} e^{-r \gamma t} \right)  \frac{\partial}{\partial V} \right>. \nonumber
\end{equation}
We find the following invariants of the subgroup $H_5$

\begin{eqnarray}\label{invH5HARAspecEXP}
inv_1 &=& z = \frac{l}{h}, ~~~ inv_2 = \tau =  r t - \log h,  \\
inv_3 &=& W(z,\tau) =  V \mp r t + \frac{1-\gamma}{r \gamma^2} e^{- r \gamma t}. \label{depvar:H5HARAspecEXP}
\end{eqnarray}

Substituting the invariants of $H_5$ (\ref{invH5HARAspecEXP}) as independent variables $z, \tau$ and the invariant (\ref{depvar:H5HARAspecEXP}) as the dependent variable $W(z, \tau)$ into (\ref{eq:HARAexpGamma}) we obtain a two dimensional PDE

 \begin{eqnarray}\label{eq:HARAspecH5}
		&&r W_{\tau}(z, \tau) \pm r + \frac{1}{2} \eta^2 \left( 2z W_z (z, \tau) + z^2 W_{zz} (z, \tau) \right) + (rz + \delta) W_z (z, \tau) - (\mu - \delta) zW_z (z, \tau)  \nonumber\\
&-& \frac{(\alpha - r)^2 W_z^2(z, \tau) - 2 (\alpha - r) \eta \rho W_z(z,\tau) (W_z (z, \tau) + z W_{zz} (z, \tau))}{2 \sigma^2 W_{zz} (z, \tau)}  \\
&-& \frac{\eta^2 \rho^2 \sigma^2 (W_z (z, \tau) + z W_{zz} (z, \tau))^2}{2 \sigma^2 W_{zz} (z, \tau)} + \frac{(1-\gamma)^2}{\gamma} e^{-\frac{\gamma}{1-\gamma}\tau} W_z^{-\frac{\gamma}{1-\gamma}}  = 0, ~~~ W(z, \tau) \underset{\tau \to \infty}{\longrightarrow} 0.\nonumber
	\end{eqnarray}
Here the investment $\pi(z, \tau ,h)$ and consumption $c(z, \tau, h)$ look as follows in terms of the value function W
\begin{eqnarray}
\label{pi:HARAexpGH3}
		 \pi (z, \tau, h) &=&  h \left( \frac{\eta \rho}{\sigma} z +\frac{\eta \rho \sigma - \alpha + r}{\sigma^2} \frac{W_z(z, \tau)}{W_{zz}(z, \tau)} \right),  \label{pi:HARAexpGH5} \\
		 c(z, \tau, h) &=& h (1-\gamma) W_z(z,\tau)^{-\frac{1}{1-\gamma}}e^{-\frac{\gamma}{1-\gamma} \tau}.  \label{c:HARAexpGH5}		
\end{eqnarray}

This equation for the new dependent variable $W(z, \tau)$ differs from (\ref{eq:HARAspecH3}) only by the term $r$ as well as the corresponding invariant (\ref{depvar:H5HARAspecEXP}) differs from (\ref{depvar:H3HARAspecEXP}) by the linear term $rt$.

\textbf{Case} $H_6(h_6)$. This subalgebra is spanned by
\begin{equation}
h_6 = <e_3 + \omega e_4> = \left< \omega l\frac{\partial}{\partial l} + \omega h\frac{\partial}{\partial h} + \frac{\omega-1}{r} \frac{\partial}{\partial t} + \left( \gamma V + \omega\frac{1-\gamma}{r \gamma} e^{-r \gamma t} \right)  \frac{\partial}{\partial V} \right>. \nonumber
\end{equation}
If $\omega = 1$ that case is equivalent to (\ref{invH4HARAEXPspec}) that we have regarded earlier, the only difference is that we need to take into consideration that now $\kappa = r \gamma$. For all the values of $\omega$ we find the following invariants of the subgroup $H_6$ solving the corresponding characteristic system

\begin{eqnarray}\label{invH6HARAspecEXP}
inv_1 &=& z = \frac{l}{h},~~~  inv_2 = \tau =  \frac{r \omega}{\omega-1}t - \log h,  \\
inv_3 &=& W(z,\tau) =  V e^{-\frac{r\gamma}{\omega-1}t} + \frac{1 -\gamma}{r\gamma^2} e^{- \frac{\omega r\gamma}{\omega-1}t}.\label{depvar:H6HARAspecEXP}
\end{eqnarray}

We use the invariants of $H_6$ (\ref{invH6HARAspecEXP}) as independent variables $z, \tau$ and the invariant (\ref{depvar:H6HARAspecEXP}) as the dependent variable $W(z, \tau)$ in (\ref{eq:HARAexpGamma}) and obtain a two dimensional equation

 \begin{eqnarray}\label{eq:HARAspecH6}
 &&\frac{r\gamma}{\omega-1}W(z, \tau) + \frac{r \omega}{\omega-1} W_\tau  (z, \tau) + \frac{1}{2} \eta^2 \left( 2z W_z (z, \tau) + z^2 W_{zz} (z, \tau) \right) \nonumber \\
&+& (rz + \delta) W_z (z, \tau) - (\mu - \delta) z W_z (z, \tau)  \\
&-& \frac{(\alpha - r)^2 W_z^2(z, \tau) - 2 (\alpha - r) \eta \rho W_z (z, \tau) (W_z (z, \tau) + z W_{zz} (z, \tau))}{2 \sigma^2 W_{zz} (z, \tau)}\nonumber \\
&+&\frac{\eta^2 \rho^2 \sigma^2 (W_z (z, \tau) + z W_{zz} (z, \tau))^2}{2 \sigma^2 W_{zz} (z, \tau)} + \frac{(1-\gamma)^2}{\gamma} e^{-\frac{\gamma}{1-\gamma} \tau} W_z^{-\frac{\gamma}{1-\gamma}}(z,\tau) = 0, ~~~ W(z, \tau) \underset{\tau \to \infty}{\longrightarrow} 0. \nonumber
\end{eqnarray}
Here the investment $\pi(z, \tau ,h)$ and consumption $c(z, \tau, h)$ look as follows in terms of the value function W
\begin{eqnarray}
		 \pi (z, \tau ,h) &=& h \left(\frac{\eta \rho}{\sigma} z + \frac{\eta \rho \sigma - \alpha + r}{\sigma^2} \frac{W_z (z, \tau)}{W_{zz}(z, \tau)} \right), \label{pi:HARAexpGH6} \\
		 c(z, \tau, h) &=&  h (1-\gamma)  e^{\frac{-\gamma}{1-\gamma} \tau} W_z^{-\frac{1}{1-\gamma}} (z, \tau).  \label{c:HARAexpGH6}
\end{eqnarray}
\textbf{Case} $H_7(h_7)$. This subalgebra is spanned by
\begin{equation}
h_7 = < e_1 \pm e_2 + \omega e_4 > = \left< \left( e^{rt} + \omega l \right)\frac{\partial}{\partial l} + \omega h\frac{\partial}{\partial h} + \frac{\omega}{r} \frac{\partial}{\partial t} + \left( \pm V + \omega \frac{1-\gamma}{r \gamma} e^{-r \gamma t} \right)  \frac{\partial}{\partial V} \right>. \nonumber
\end{equation}
Let us note here that if $\omega = 0$ this case is equivalent to the case (\ref{charsystemH4}) under a condition $\kappa = r \gamma$ that we have regarded before. If $\omega \neq 0$ we can find the following invariants of the subgroup $H_7$

\begin{eqnarray}\label{invH7HARAspecEXP}
inv_1 &=& x = l e^{-rt} - \frac{r}{\omega}t, ~~~ inv_2 = y =  h e^{-rt},  \\
inv_3 &=& W(x,y) =  V \mp \frac{r}{\omega}t + \frac{1 -\gamma}{r\gamma^2} e^{- r\gamma t}.\label{depvar:H7HARAspecEXP}
\end{eqnarray}

Substituting the invariants of $H_7$ (\ref{invH7HARAspecEXP}) as the independent variables $x$ and $y$ and the invariant (\ref{depvar:H7HARAspecEXP}) as the dependent variable $W(x,y)$ into (\ref{eq:HARAexpGamma}) we obtain a two dimensional PDE

 \begin{eqnarray}\label{eq:HARAspecH7}
&\pm& \frac{r}{\omega} - \frac{r}{\omega} W_x(x, y)  + \frac{1}{2}\eta^2 y^2 W_{yy} (x, y)  + \delta y W_x (x, y) + (\mu -\delta)y W_{y}(z, y)   \\
&-& \frac{(\alpha - r)^2 W_x^2(x, y)  - 2(\alpha - r) \eta \rho y W_x (x, y) W_{x y}(x, y)  + \eta^2 \rho^2 \sigma^2 y^2 W_{x y}^2(x, y) }{2\sigma^2W_{xx}(x, y) }\nonumber\\
&+& \frac{(1-\gamma)^2}{\gamma} W_x^{-\frac{\gamma}{1-\gamma}} (x, y) = 0, ~~~W(x,0) \underset{x \to \mp \infty}{\longrightarrow} 0.\nonumber
\end{eqnarray}

Here the investment $\pi(x, y ,h)$ and consumption $c(x, y, h)$ look as follows in terms of the value function W
\begin{eqnarray}
		\pi (x, y ,h) &=& - h\frac{\eta \rho \sigma W_{xy}(x, y) + (\alpha - r) y^{-1} W_x (x, y)}{\sigma^2 W_{xx}(x, y)}, \label{pi:HARAexpGH7} \\
		c(x, y, h) &=& h (1-\gamma) y^{-1} W_x^{-\frac{1}{1-\gamma}}(x,y).   \label{c:HARAexpGH7}
\end{eqnarray}

\subsubsection{Two-dimensional subalgebras of $L^{HARA}_4$}

In this section we go further and using Table \ref{TableOptSystL4HARA2} find the deeper reductions that can reduce PDE (\ref{eq:HARAexpGamma}) to an ordinary differential equation. At first let us note that all two dimensional subalgebras listed in Table \ref{TableOptSystL4HARA} but for $h_{11}$ do not give a reduction of the original equation (\ref{eq:HARAexpGamma}) to an ODE, so we will not regard them in detail.

\textbf{Case} $H_{11}(h_{11})$. The first two dimensional sub algebra listed in Table \ref{TableOptSystL4HARA} is subalgebra $h_{11} = < e_3, e_4 >$ spanned by two generators defined as follows
\begin{eqnarray} \label{gen:H11}
e_3 &=& - \frac{1}{ r} \frac{\partial}{\partial t} + \gamma V \frac{\partial}{\partial V}  \\
e_4 &=& l\frac{\partial}{\partial l} + h\frac{\partial}{\partial h} + \frac{1}{r} \frac{\partial}{\partial t} + \frac{1-\gamma}{r \gamma} e^{-r \gamma t}  \frac{\partial}{\partial V}. \nonumber
\end{eqnarray}
The two dimensional subalgebra is spanned by two generators. We should find a simultaneous solution to both characteristic systems. We can use our previous knowledge and reformulate one of the characteristic systems in terms of invariant variables of the other one.  This could be done in the following way. We have found a general form of the invariants of $e_4$ in (\ref{invH3HARAspecEXP}).  Let us list them here again for the convenience of the reader
\begin{eqnarray}
inv_1 &=& z = \frac{l}{h}, ~~~ inv_2 = \tau =  r t - \log h, \nonumber \\
inv_3 &=& W(z,\tau) =  V + \frac{1 -\gamma}{r\gamma^2} e^{- r \gamma t}. \nonumber
\end{eqnarray}
If we rewrite $e_3$ (\ref{gen:H11}) in terms of these three invariants $z, \tau$ and $W$ as new independent and dependent variables correspondingly, we obtain
\begin{equation}
e_3 = - \frac{\partial}{\partial \tau} + \gamma W \frac{\partial}{\partial W}.
\end{equation}
Solving a corresponding characteristic system $\frac{d \tau}{- 1} = \frac{d W}{\gamma W}$ we obtain a new invariant
\begin{equation}
inv_{e_3} = Y(z) =  W(z, \tau) e^{\gamma \tau}.
\end{equation}
This way we obtain an invariant solution under the action of two parameter subgroup $H_{11}$. It means that we can take $Y(z)$ as a new dependent variable in (\ref{eq:HARAspecH3}) and $z$ as a new independent one. Substituting these invariants into PDE (\ref{eq:HARAspecH3}) we obtain a new ODE
\begin{eqnarray}\label{eq:HARAspecH11}
&-&r \gamma Y  (z) + \frac{1}{2} \eta^2 \left( 2z Y_z (z) + z^2 Y_{zz} (z) \right) + (r z + \delta) Y_z (z) - (\mu - \delta) z Y_z (z)  \nonumber\\
&-& \frac{(\alpha - r)^2 Y_z^2(z) - 2 (\alpha - r) \eta \rho Y_z (z) (Y_z (z) + z Y_{zz} (z)) + \eta^2 \rho^2 \sigma^2 (Y_z (z) - z Y_{zz} (z))^2}{2 \sigma^2 Y_{zz} (z)} \nonumber \\
&+& \frac{(1-\gamma)^2}{\gamma} Y_z^{-\frac{\gamma}{1-\gamma}} (z)  = 0.
\end{eqnarray}
Naturally, the investment $\pi(z, \tau ,h)$ and consumption $c(z, \tau , h)$ in terms of $Y(z)$ look now as follows

\begin{eqnarray}
		 \pi (z, \tau ,h) &=& h \left( \frac{\eta \rho}{\sigma} z +\frac{\eta \rho \sigma - \alpha + r}{\sigma^2} \frac{Y_z (z)}{Y_{zz}(z) } \right), \label{pi:HARAspecGH11} \\
		 c(z, \tau, h) &=&  h (1-\gamma) Y_z(z)^{-\frac{1}{1-\gamma}}.  \label{c:HARAspecGH11}
		 \end{eqnarray}
In terms of original variables $l, h, t$ and $V$ the substitution looks as follows
\begin{eqnarray}
z &=& \frac{l}{h}, ~~~  \tau =  r t - \log h, \nonumber  \\
Y(z) &=& \left( V e^{\gamma r t} + \frac{1 -\gamma}{r\gamma^2} \right)  h^{-\gamma}. \nonumber
\end{eqnarray}
This substitution is equivalent to the substitution (\ref{2subHARAH8}) under the condition $\kappa = r \gamma$.
It is also important to note that if we obtain a solution $Y(z)$ for (\ref{eq:HARAspecH11}) we obtain the value function that in terms of original variables looks like
$$
V(t, l ,h) = e^{-\kappa t} h^{\gamma} Y(l/h) - \frac{1-\gamma}{\gamma \kappa} e^{- \kappa t},
$$
and the condition $V(l, h, t) \underset{t \to \infty}{\longrightarrow} 0$ is satisfied.

\section{Reductions for a general liquidation time distribution and log-utility function}

In this chapter we discuss the complete set of possible symmetry reductions of three dimensional PDE (\ref{maingeneralLOG}) arising in the case of log-utility function and a general liquidation time distribution. We also regard a special case of an exponentially distributed liquidation time in the Section \ref{sec:logexp}. The logarithmic utility could be regarded as a special case of HARA-utility, as we have mentioned earlier (\ref{HARAlimLOG}), moreover we see that $L^{HARA}_{3,4} \underset{\gamma \to 0}{\longrightarrow} L^{LOG}_{3,4}$. Yet logarithmic utility is widely used in financial mathematics and therefore deserves our attention.

\subsection{The case of the general liquidation time distribution}

Analogously to the previous sections we look for the optimal system of subalgebras for the Lie algebra $L^{LOG}_3$ (\ref{symalloggeneral}) admitted by the equation (\ref{maingeneralLOG}). It is necessary to study an optimal system of Lie subalgebras to be able to describe all non equivalent symmetry reductions and not loose any of them. We present here for a convenience of the reader the main equation
\begin{eqnarray}\label{eq:generalLOG}
		 && V_t(t, l, h) +  \frac{1}{2}\eta^2h^2V_{hh} (t, l, h) + (rl + \delta h)V_l (t, l, h) + (\mu - \delta) hV_h (t, l, h)  \nonumber \\
		 &-& \frac{(\alpha - r)^2 V_{l}^2(t,l,h)+2(\alpha - r)\eta \rho h V_{l}(t,l,h) V_{lh}(t,l,h) + \eta^2 \rho^2 \sigma^2 h^2 {V_{lh}}^2 (t,l,h)}{2 \sigma^2 V_{ll}(t,l,h)} \nonumber \\
 		&-& \overline{\Phi}(t)\left( \log V_l (t, l, h)- \log \overline{\Phi}(t) + 1\right)=0, ~~~V \underset{t \to \infty}{\longrightarrow} 0.
	\end{eqnarray}
Here the investment $\pi(t, l ,h)$ and consumption $c(t, l, h)$ look as follows in terms of the value function V
\begin{eqnarray}
		 \pi (t, l ,h) &=& - \frac{\eta \rho \sigma h V_{lh}(t, l, h) + (\alpha - r) V_l (t, l, h)}{\sigma^2 V_{ll}(t,l,h)}, \label{pi:LOG} \\
		 c(t, l, h) &=& \frac{ \overline{\Phi}(t)}{V_l (t, l, h)}.  \label{c:LOG}
\end{eqnarray}
We list all symmetry reductions that can reduce the dimension of the problem by one or two and study them closely in the next Section. We demonstrate how the problem (\ref{eq:generalLOG}) transforms after every substitution.

\subsubsection{System of optimal subalgebras for $L^{LOG}_3$ and possible invariant reductions}

At first let us reassign the basis of $L^{LOG}_3$ (\ref{symalloggeneral}) so that it is possible to use the  classification \cite{patera&wintern} in a convenient way. This can be done in the following way ${\mathbf U}_1 = - e_3, {\mathbf U}_2 =  e_2, {\mathbf U}_3 = - e_1$. The basis is now defined as
\begin{equation} \label{basisL3LOG}
	e_1 =  - l \frac{\partial}{\partial l} - h \frac{\partial}{\partial h} + \int \overline{\Phi}(t)  dt \frac{\partial}{\partial V},  ~~~ e_2 = e^{rt} \frac{\partial}{\partial l}, ~~~ e_3 = - \frac{\partial}{\partial V}.
\end{equation}

The new basis has just one non zero commutation relation

\begin{equation} \label{comL3LOG}
	[e_1,e_2] = e_2.
\end{equation}

 Now we can see that $L^{LOG}_3$ corresponds to $A_1 \bigoplus A_2$  case, in the notation of \cite{patera&wintern}. The system of optimal subalgebras is given in Table \ref{TableOptSystL3LOG}.

\begin{table}[h]
\begin{center}
\begin{tabular}{|l|l|}
         \hline
         Dimension of & $~~~~~~~~~~~~~~~~~~~~~$System of optimal subalgebras of algebra $L^{LOG}_3$ \\
         the subalgebra & \\
         \hline

         1                           & $h_1=\left< e_1 \cos \beta + e_3 \sin  \beta  \right>, ~ h_2= \left< e_2 \pm e_3 \right>, ~h_3=\left< e_2 \right>$\\
  \hline
         2                           & $h_4=\left < e_1, e_3 \right>, ~h_5=\left <e_2, e_3 \right>,~ h_6=\left <e_1 + \omega e_3 , e_2\right>$\\

         \hline
\end{tabular}
\caption{ The optimal system of one- and two-dimensional
subalgebras of $L^{LOG}_3$ (\ref{symalharageneral}), where $\omega$ and $ \beta$ are parameters such that $-\infty < \omega < \infty$ and $0 \leq  \beta < \pi $. \label{TableOptSystL3LOG}}
\end{center}
\end{table}

\subsubsection{One-dimensional subalgebras of $L^{LOG}_3$ and corresponding reduction}

Let us now look at one dimensional subalgebras of $L^{LOG}_3$ one by one to find all non equivalent reductions of (\ref{eq:generalLOG}).

\textbf{Case} $H_1(h_1)$. This subalgebra is spanned by the generator
\begin{equation}
h_1 = < e_1 \cos \beta + e_3 \sin  \beta  > =  \left< \cos \beta l \frac{\partial}{\partial l} + \cos \beta h \frac{\partial}{\partial h} +\left( \sin  \beta - \cos \beta \int \overline{\Phi}(t)  dt \right)\frac{\partial}{\partial V} \right>. \nonumber
\end{equation}
Solving a corresponding characteristic system for the invariants of $H_1$ we obtain independent invariants
\begin{eqnarray}\label{invH1LOG}
inv_1 &=& z = \frac{l}{h}, ~~~  inv_2 = t \\
inv_3 &=& W(z,t) = V + \log h \left( \tan  \beta + \int \overline{\Phi}(t)  dt \right). \label{depvar:H1LOG}
\end{eqnarray}
Substituting (\ref{invH1LOG}) as new independent variables $t, z$ and (\ref{depvar:H1LOG}) as a new dependent variable $W(z,t)$ into (\ref{maingeneralLOG}) we get

\begin{eqnarray}\label{2dimLOGgen}
&&W_t(z,t) + \frac{1}{2} \eta^2 \left( 2z W_z (z, t) + z^2 W_{zz} (z, t) \right) +(r z + \delta) W_z (z, t) - (\mu - \delta) z W_z (z, t)  \nonumber\\
&-& \frac{(\alpha - r)^2 W_z^2(z, t) - 2 (\alpha - r) \eta \rho W_z (z, t) (W_z (z, t) + z W_{zz} (z, t)) + \eta^2 \rho^2 \sigma^2 (W_z (z, t) - z W_{zz} (z, t))^2}{2 \sigma^2 W_{zz} (z, t)} \nonumber \\
&-&  \overline{\Phi}(t) \log W_z(z,t) - \left(\frac{1}{2}\eta^2 -\mu + \delta \right) \int \overline{\Phi}(t)  dt +  \overline{\Phi}(t) (\log \overline{\Phi}(t) - 1)\nonumber \\
&-& \left(\frac{1}{2}\eta^2 -\mu + \delta \right) \tan  \beta = 0. \nonumber
\end{eqnarray}
Indeed the condition $V \underset{t \to \infty}{\longrightarrow} 0$ holds only for the situation when $\beta = 0$. When $\beta=0$, the investment $\pi(z, t , h)$ and consumption $c(z, t, h)$ look as follows in terms of the value function W
\begin{eqnarray}
		\pi (z, t ,h) &=& h \left( \frac{\eta \rho}{\sigma} z + \frac{\eta \rho \sigma - \alpha + r}{\sigma^2} \frac{W_z (z, t)}{W_{zz}(z, t)} \right), \label{pi:LOGH1} \\
		 c(z, t, h) &=& h \frac{ \overline{\Phi}(t)}{W_z (z, t)}.  \label{c:LOGH1}
\end{eqnarray}
\textbf{Cases} $H_2$ spanned by $h_2 = \left< e^{rt} \frac{\partial}{\partial l}\right>$ and $H_3(h_3)$ spanned by $h_3 = \left< \frac{\partial}{\partial V}\right>$ were in principle discussed in Section \ref{sec:LHARA3} hence we do not speak about them in detail.
We can see that for a general liquidation time distribution we have only one meaningful Lie-type reduction of the equation (\ref{eq:generalLOG}). This does not mean that the problem can not be simplifies or solved by the means of numeric or quantitive analysis though.

\subsection{A special case of an exponential liquidation time distribution and log utility function}\label{sec:logexp}
In Theorem \ref{MTlog} we have shown that the case of an exponential liquidation time distribution is a special one and the equation (\ref{maingeneralLOG}) admits an extended Lie algebra. There are four Lie-symmetries in this case, describe in (\ref{symalloggeneral}) and (\ref{exp4symlog}). We would like to pay some special attention to the case of a logarithmic utility since this particular case is broadly regarded in the literature. The equation (\ref{eq:generalLOG}) we study in this section now, when we insert $\overline{\Phi}(t) = e^{- \kappa t}$, looks as follows

\begin{eqnarray}\label{eq:LOGexp}
		 && V_t(t, l, h) +  \frac{1}{2}\eta^2h^2V_{hh} (t, l, h) + (rl + \delta h)V_l (t, l, h) + (\mu - \delta) hV_h (t, l, h)  \nonumber \\
		 &-& \frac{(\alpha - r)^2 V_{l}^2(t,l,h)+2(\alpha - r)\eta \rho h V_{l}(t,l,h) V_{lh}(t,l,h) + \eta^2 \rho^2 \sigma^2 h^2 {V_{lh}}^2 (t,l,h)}{2 \sigma^2 V_{ll}(t,l,h)} \nonumber \\
 		&-&e^{-\kappa t}\left( \log V_l  (t, l, h) + \kappa t + 1\right)=0, V \underset{t \to \infty}{\longrightarrow} 0.
	\end{eqnarray}
Here the investment $\pi(t, l ,h)$ and consumption $c(t, l, h)$ look as follows in terms of the value function V
\begin{eqnarray}
		 \pi (t, l ,h) &=& - \frac{\eta \rho \sigma h V_{lh}(t, l, h) + (\alpha - r) V_l (t, l, h)}{\sigma^2 V_{ll}(t,l,h)}, \label{pi:LOGexp} \\
		 c(t, l, h) &=& \frac{ e^{-\kappa t}}{V_l (t, l, h)}.  \label{c:LOGexp}
\end{eqnarray}
\subsubsection{Study of non equivalent reductions with the system of optimal subalgebras}
As in previous cases we change the basis of $L^{LOG}_4$ to use the convenient classification provided in \cite{patera&wintern}.
Let us at first transform the basis as follows $$e_1 = \frac{r \mathbf U_3 + \mathbf U_4}{\kappa},~~~ e_2 = \mathbf U_1, ~~~e_3 = - \mathbf U_3, ~~~e_4 = \mathbf U_2.$$

Now the generators of the new basis of $L_4 = < e_1, e_2, e_3, e_4>$ look like this

\begin{eqnarray}
		&& e_1 =  \frac{r}{\kappa} l\frac{\partial}{\partial l} +  \frac{r}{\kappa} h\frac{\partial}{\partial h} + \frac{1}{\kappa} \frac{\partial}{\partial t} - \left( V - \frac{r}{\kappa^2} e^{-\kappa t} \right) \frac{\partial}{\partial V}, \nonumber\\
		&& e_2 = \frac{\partial}{\partial V},\nonumber \\
		&& e_3 = - l \frac{\partial}{\partial l} -  h\frac{\partial}{\partial h}  - \frac{1}{\kappa} e^{-\kappa t} \frac{\partial}{\partial V}, \nonumber\\
		&& e_4= e^{rt} \frac{\partial}{\partial l}.\nonumber
\end{eqnarray}
In this basis there are only two non-zero commutation relations on $L^{LOG}_4$
\begin{equation} \label{comL4LOG}
	[e_1,e_2] = e_2,~~~  [e_3, e_4]= e_4.
\end{equation}

 Now we can see that $L^{LOG}_4$ corresponds to $2A_{2}$, in the notation of \cite{patera&wintern}.
 \begin{remark} If $\gamma \to 0$ for $\kappa \neq r \gamma$ in (\ref{basisWinternL4HARA}) then the basis of $L^{HARA}_4$ transforms into the basis of $L^{LOG}_4$ in this case.
 \end{remark}
 The system of optimal subalgebras of $L^{LOG}_4$ is listed in Table \ref{TableOptSystL4LOG}.
\begin{table}[h]
\begin{center}
\begin{tabular}{|l|l|}
         \hline
         Dimension of & $~~~~~~~~~~~~~~~~~~~~~$System of optimal subalgebras of algebra $L^{LOG}_4$ \\
         the subalgebra & \\
         \hline

         1                       & $h_1=\left< e_2 \right>, ~ h_2= \left< e_3 \right>, ~h_3=\left< e_4 \right>, ~h_4=\left< ~e_1 + \omega e_3 \right>, ~h_5=\left< ~e_1 \pm e_4 \right>,$ \\
         			& $h_6=\left< ~e_2 \pm e_4 \right>, ~h_7=\left< ~e_2 \pm e_3 \right> $\\
  \hline
         2                       & $h_ 8=\left < e_1, e_3 \right>, ~h_9=\left <e_1, e_4\right>,~ h_{10}=\left <e_2,e_3\right>,~ h_{11}=\left <e_2,e_4\right>,$\\
         			& $~ h_{12}=\left <e_1+\omega e_3,e_2\right>, ~ h_{13}=\left <e_3 + \omega e_1,e_4\right>, ~ h_{14}=\left <e_1 \pm e_4,e_2\right>,$\\
				& $~ h_{15}=\left <e_3 \pm e_2,e_4\right>, ~ h_{16}=\left <e_1 + e_3,e_2 \pm e_4\right>$\\

         \hline
          3                      & $h_{17}=\left < e_1, e_3, e_2 \right>, h_{18}=\left < e_1, e_4, e_2 \right>, h_{19}=\left < e_1, e_3, e_4 \right>, h_{20}=\left < e_2, e_3, e_4 \right>, $\\
				& $h_{21}=\left < e_1 \pm e_3, e_2, e_4 \right>, h_{22}=\left < e_1 + \omega e_3, e_2, e_4 \right>$\\
         \hline
\end{tabular}
\caption{ The optimal system of one-, two- and three- dimensional
subalgebras of $L^{LOG}_4$,  where $\omega$ is a parameter such that $-\infty < \omega < \infty$. \label{TableOptSystL4LOG}}
\end{center}
\end{table}

\subsubsection{One-dimensional subalgebras of $L^{LOG}_4$ and related invariant reductions}
Now we are going to study all invariant reductions of the problem (\ref{eq:LOGexp}). Let us first note that the subgroups $H_1$, $H_3$ and $H_6$, spanned by the generators $h_1 = \left< \frac{\partial }{\partial V} \right>$, $h_3 = \left< e^{rt} \frac{\partial }{\partial l} \right>$ and $h_6 = \left< \frac{\partial }{\partial V} \pm  e^{rt} \frac{\partial }{\partial l} \right>$ correspondingly, do not give us any interesting reductions so we omit the detailed study of these cases here. We start with a first interesting and non-trivial case.

\textbf{Case} $H_2(h_2)$.  The sub algebra is spanned by the generator
\begin{equation}
h_2 = < e_3 >= \left< - l \frac{\partial}{\partial l} -  h\frac{\partial}{\partial h}  - \frac{1}{\kappa} e^{-\kappa t} \frac{\partial}{\partial V} \right>. \nonumber
\end{equation}
To find all invariants of the subgroup $H_2$ we solve the corresponding characteristic system of equations and obtain
\begin{eqnarray}\label{invH2LOG}
inv_1 &=& z = \frac{l}{h}, ~~~ inv_2 = t \nonumber \\
inv_3 &=& W(z,t) = \kappa e^{\kappa t} V - \log h . \label{depvar:H2LOG}
\end{eqnarray}

Substituting two invariants of $H_2$ (\ref{invH2LOG}) as the new independent variables $z, t$ and (\ref{depvar:H2LOG}) as the dependent variable $W(z,t)$ into (\ref{eq:LOGexp}) we get

\begin{eqnarray}\label{2dimLOGH2}
W_t(z,t) &-&\kappa W (z,t) + \frac{1}{2} \eta^2 \left( 2z W_z (z, t) + z^2 W_{zz} (z, t) \right) +(r z + \delta) W_z (z, t) - (\mu - \delta) z W_z (z, t)  \nonumber\\
&-& \frac{(\alpha - r)^2 W_z^2(z, t) - 2 (\alpha - r) \eta \rho W_z (z, t) (W_z (z, t) + z W_{zz} (z, t))}{2 \sigma^2 W_{zz} (z, t)} \nonumber \\
&-&\frac{\eta^2 \rho^2 \sigma^2 (W_z (z, t) + z W_{zz} (z, t))^2}{2 \sigma^2 W_{zz} (z, t)} -  \kappa \log W_z (z,t) \nonumber \\
&-& \left(\frac{1}{2}\eta^2 -\mu + \delta \right) +  \kappa (\log \kappa - 1)=0, ~~~ W \underset{t \to \infty}{\longrightarrow} 0. \nonumber
\end{eqnarray}
Here the investment $\pi(z, t ,h)$ and consumption $c(z, t, h)$ look as follows in terms of the  function W
\begin{eqnarray}
		 \pi (z, t ,h) &=& h \left( \frac{\eta \rho}{\sigma} z  + \frac{\eta \rho \sigma - \alpha + r}{\sigma^2} \frac{W_z (z, t)}{W_{zz}(z, t)} \right), \label{pi:LOGexpH2} \\
		 c(z, t, h) &=& h \frac{\kappa}{W_z (z, t)}. \label{c:LOGexpH2}
\end{eqnarray}
\textbf{Case} $H_4(h_4)$. This subalgebra is spanned by
\begin{equation}
h_4 = <e_1 + \omega e_3 > = \left< \left( \frac{r}{\kappa} -\omega \right) l\frac{\partial}{\partial l} + \left( \frac{r}{\kappa} -\omega \right) h\frac{\partial}{\partial h} + \frac{1}{\kappa} \frac{\partial}{\partial t} - \left( V - \frac{r - \omega \kappa}{\kappa^2} e^{-\kappa t} \right) \frac{\partial}{\partial V} \right>. \nonumber
\end{equation}
We need to regard two special cases $\omega = r/\kappa$ and $\omega \neq r/\kappa$ here. If $\omega = r/\kappa$ this case $h_4$ is spanned by a generator
\begin{equation}
h_4=<e_1 + \frac{r}{\kappa} e_3>=\frac{1}{\kappa} \frac{\partial}{\partial t} - V \frac{\partial}{\partial V}. \nonumber
\end{equation}
The invariants for this case are as follows
\begin{eqnarray}\label{invH4LOGspec}
inv_1 &=& l , ~~~ inv_2 = h ,\\
inv_3 &=& W(l,h) = V e^{\kappa t} \label{depvar:H4LOGspec}
\end{eqnarray}
Using two invariants (\ref{invH4LOGspec}) as the new independent variables and (\ref{depvar:H4LOGspec}) as the dependent variable in (\ref{eq:LOGexp}) we obtain a two dimensional PDE
\begin{eqnarray}\label{eq:H4LOGspec}
		 &-\kappa& W(l, h) +  \frac{1}{2}\eta^2h^2W_{hh} (l, h) + (rl + \delta h)W_l (l, h) + (\mu - \delta) hW_h (l, h)  \nonumber \\
		 &-& \frac{(\alpha - r)^2 W_{l}^2(l,h)+2(\alpha - r)\eta \rho h W_{l}(l,h) W_{lh}(l,h) + \eta^2 \rho^2 \sigma^2 h^2 {W_{lh}}^2 (l,h)}{2 \sigma^2 W_{ll}(l,h)} \nonumber \\
 		&-&\left( \log W_l  (l, h) + 1\right)=0,~~~ W \underset{t \to \infty}{\longrightarrow} 0.
	\end{eqnarray}
We see that in this case the value function $V(l,h,t) = e^{-\kappa t} W(l, h)$ and the complete dependence on $t$ is described just by the factor $e^{- \kappa t}$
Here the investment $\pi(l ,h)$ and consumption $c(l, h)$ look as follows in terms of the function W
\begin{eqnarray}
		 \pi (l ,h) &=& - \frac{\eta \rho \sigma h W_{lh}(l, h) + (\alpha - r) W_l (l, h)}{\sigma^2 W_{ll}(t,l,h)}, \label{pi:LOGexpH4spec} \\
		 c(l, h) &=& h \frac{\kappa}{W_z (l, h)}.  \label{c:LOGexpH4spec}
\end{eqnarray}
To find the invariants of $H_4$  when $\omega \neq r/ \kappa$ we can move according to a standard procedure. We obtain three independent invariants using a corresponding characteristic system
\begin{eqnarray}\label{invH4LOG}
inv_1 &=& z = \frac{l}{h}, ~~~ inv_2 = \tau = (r - \kappa \omega)t - \log h ,\\
inv_3 &=& W(z,\tau) =  e^{\kappa t} V + \left( \omega - \frac{r}{\kappa} \right)t . \label{depvar:H4LOG}
\end{eqnarray}

Analogously substituting expressions for the invariants $z$ and $\tau$ (\ref{invH4LOG}) as the new independent and $W(z, \tau)$ (\ref{depvar:H4LOG}) as the new dependent variables into (\ref{maingeneralLOG}) we get

\begin{eqnarray}\label{2dimLOGH4}
&&(r - \kappa \omega )W_{\tau} (z,\tau)-\kappa W (z,\tau) + \frac{1}{2} \eta^2 \left( 2z W_z (z, \tau) + z^2 W_{zz} (z,\tau) \right) \nonumber\\
&+&(r z + \delta) W_z (z, \tau) - (\mu - \delta) z W_z (z, \tau)  \\
&-& \frac{(\alpha - r)^2 W_z^2(z,\tau) - 2 (\alpha - r) \eta \rho W_z (z, \tau) (W_z (z, \tau) + z W_{zz} (z, \tau)) }{2 \sigma^2 W_{zz} (z,\tau)} \nonumber \\
&-&\frac{\eta^2 \rho^2 \sigma^2 (W_z (z, \tau) + z W_{zz} (z, \tau))^2}{2 \sigma^2 W_{zz} (z,\tau)} \nonumber \\
&-&   \log W_z (z,\tau) - \tau  - \omega+ \frac{r}{\kappa} -1= 0,~~~ W \underset{t \to \infty}{\longrightarrow} 0.
\end{eqnarray}
Here the investment $\pi(z, \tau ,h)$ and consumption $c(z, \tau, h)$ look as follows in terms of the value function W
\begin{eqnarray}
		 \pi (z, \tau ,h) &=& h \left(\frac{\eta \rho}{\sigma} z + \frac{\eta \rho \sigma - \alpha + r}{\sigma^2} \frac{W_z (z, \tau)}{W_{zz}(z, \tau)} \right), \label{pi:LOGexpH4} \\
		 c(z, \tau ,h) &=& h \frac{\kappa}{W_z (z, \tau)}.  \label{c:LOGexpH4}
\end{eqnarray}
\textbf{Case} $H_5(h_5)$.This subalgebra is spanned by
\begin{equation}
h_5 = <e_1 \pm e_4> = \left< \left( \frac{r}{\kappa}l \pm e^{rt}\right) \frac{\partial}{\partial l} + \frac{r}{\kappa} h\frac{\partial}{\partial h} + \frac{1}{\kappa} \frac{\partial}{\partial t} - \left( V - \frac{r}{\kappa^2} e^{-\kappa t} \right) \frac{\partial}{\partial V} \right>. \nonumber
\end{equation}
According to a standard procedure for finding the invariants of the subgroup $H_5$ we obtain three independent invariants as a solution of the characteristic system
\begin{eqnarray}\label{invH5LOG}
inv_1 &=& x = l e^{-rt} \mp \kappa t, ~~~ inv_2 = y = h e^{-rt},\\
inv_3 &=& W(x, y) = e^{\kappa t} V +  \frac{r}{\kappa}t . \label{depvar:H5LOG}
\end{eqnarray}

Substituting the new independent variables $x, y$ (\ref{invH5LOG}) and the new dependent variable $W(x, y)$ (\ref{depvar:H5LOG}) into (\ref{maingeneralLOG}) we get a two dimensional PDE

\begin{eqnarray}\label{2dimLOGH5}
&\pm& \kappa W_{x}(x, y) -\kappa W (x, y) + \frac{1}{2}\eta^2 \tau^2W_{yy} (x, y) + \delta y W_x (x, y) + (\mu - \delta) y W_{y} (x, y)  \nonumber \\
		 &-& \frac{(\alpha - r)^2 W_{x}^2(x, y)+2(\alpha - r)\eta \rho y W_{x}(x, y) W_{x y}(x, y) + \eta^2 \rho^2 \sigma^2 y^2 {W_{x y}}^2 (x, y)}{2 \sigma^2 W_{xx}(x, y)} \nonumber \\
 		&-& \log W_x(x, y)  + \frac{r}{\kappa} - 1= 0,~~~ W(x, 0) \underset{x \to \mp \infty}{\longrightarrow} 0.
\end{eqnarray}

Here the investment $\pi(x, y ,h)$ and consumption $c(x, y, h)$ look as follows in terms of the value function W
\begin{eqnarray}
		 \pi (x, y ,h) &=& - h \frac{\eta \rho \sigma W_{xy}(x, y) + (\alpha - r) y^{-1} W_x (x, y)}{\sigma^2 W_{xx}(x,y)}, \label{pi:LOGexpH5} \\
		 c(x, y, h) &=& h \frac{1}{W_x (x, y)}. \label{c:LOGexpH5}
\end{eqnarray}
\textbf{Case} $H_7(h_7)$. The last one dimensional subalgebra in the list of optimal system of subalgebras in Table \ref{TableOptSystL4LOG} is spanned by
\begin{equation}
h_7 = <e_2 \pm e_3 > = \left< l \frac{\partial}{\partial l} +  h\frac{\partial}{\partial h} + \left( \frac{1}{\kappa} e^{-\kappa t} \mp 1 \right) \frac{\partial}{\partial V} \right>.\nonumber
 \end{equation}
 According to a standard procedure we look for invariants of the subgroup $H_7$ and obtain three independent invariants
\begin{eqnarray}\label{invH7LOG}
inv_1 &=& t, ~~~ inv_2 = z  = \frac{l}{h},\\
inv_3 &=& W(z, t) =  V -  \left( \frac{1}{\kappa} e^{-\kappa t} \mp 1 \right) \log h \label{depvar:H7LOG}
\end{eqnarray}

Using the invariants (\ref{invH7LOG}) as the new independent variables $z, t$ and the invariant (\ref{depvar:H7LOG}) as the new dependent variable $W(z,t)$ and substituting them into (\ref{maingeneralLOG}) we obtain a two dimensional PDE
\begin{eqnarray}\label{2dimLOGH7shift}
 W_{t} (z, t)&+& \frac{1}{2} \eta^2 \left( 2z W_z (z, t) + z^2 W_{zz} (z, t) \right) +(r z + \delta) W_z (z, t) - (\mu - \delta) z W_z (z, t)  \nonumber\\
&-& \frac{(\alpha - r)^2 W_z^2(z, t) - 2 (\alpha - r) \eta \rho W_z (z, t) (W_z (z, t) + z W_{zz} (z, t))}{2 \sigma^2 W_{zz} (z, t)}  \\
 &-& \frac{\eta^2 \rho^2 \sigma^2 (W_z (z, t) - z W_{zz} (z, t))^2}{2 \sigma^2 W_{zz} (z, t)} - e^{-\kappa t} \left( \log W_z(z, t) - \frac{1}{2} \eta^2 - \mu + \delta \right) = 0. \nonumber
\end{eqnarray}
We list this reduction here, but need to note that this substitution means that condition (\ref{cond:main}) is not satisfied. So, the solution of (\ref{2dimLOGH7shift}) would not be a solution of the original problem, that is why we do not demonstrate how  $\pi(z, t ,h)$ and $c(z, t, h)$ look  in this case.

Totally there are four meaningful reductions for the case of logarithmic utility and exponential liquidation time distribution.

\subsubsection{Two-dimensional subalgebras of $L^{LOG}_4$}
In this section we go further and using Table \ref{TableOptSystL4LOG} find the deeper reductions that can reduce PDE (\ref{eq:LOGexp}) to an ordinary differential equation. Two achieve this goal we need to study two parameter subalgebras.

\textbf{Case} $H_8(h_8)$. The first two dimensional subalgebra listed in Table \ref{TableOptSystL4LOG} is subalgebra $h_8 = < e_1, e_3 >$ spanned by two generator defined as follows
\begin{eqnarray}
e_1 &=&  \frac{r}{\kappa} l\frac{\partial}{\partial l} +  \frac{r}{\kappa} h\frac{\partial}{\partial h} + \frac{1}{\kappa} \frac{\partial}{\partial t} - \left( V - \frac{r}{\kappa^2} e^{-\kappa t} \right) \frac{\partial}{\partial V}, \nonumber\\
e_3 &=& - l \frac{\partial}{\partial l} -  h\frac{\partial}{\partial h}  - \frac{1}{\kappa} e^{-\kappa t} \frac{\partial}{\partial V}, \nonumber\\
\nonumber
\end{eqnarray}
Both of these generators were studied before, we can use our previous knowledge and reformulate one of the characteristic systems in terms of invariant variables of the other one.  This could be done in the following way. We have found a general form of the invariants of $e_1$. Indeed if we assume that $\omega = 0$ in (\ref{invH4LOG}) we get the corresponding invariants
\begin{eqnarray}
inv_1 &=& z = \frac{l}{h}, ~~~ inv_2 = \tau = rt - \log h , \nonumber \\
inv_3 &=& W(z,\tau) =  e^{\kappa t} V - \frac{r}{\kappa}t . \nonumber
\end{eqnarray}
If we rewrite the second generator $e_3$ of subalgebra $h_8$ in terms of these three invariants $z, \tau$ and $W$ as new independent and dependent variables correspondingly, we obtain
\begin{equation}
e_3 =\frac{\partial}{\partial \tau} - \frac{1}{\kappa} \frac{\partial}{\partial W}.
\end{equation}
Solving a corresponding characteristic system $\frac{d \tau}{1} = \frac{d W}{- 1/\kappa}$ we obtain a new common invariant
\begin{equation}
inv_{e_3} = Y(z) =  \kappa W(z, \tau) + \tau.
\end{equation}
This way we obtain an invariant solution under the action of two parameter subgroup $H_8$. It means that we can take $Y(z)$ as a new dependent variable in (\ref{eq:HARAH2}) and $z$ as a new independent one. Substituting these invariants into PDE (\ref{eq:HARAH2}) with $\omega = 0$ we obtain a new ODE
\begin{eqnarray}\label{eq:HARAH8}
&&Y(z) + \frac{1}{2} \eta^2 \left( 2z Y_z (z) + z^2 Y_{zz} (z) \right) +(r z + \delta) Y_z (z) - (\mu - \delta) z Y_z (z)  \nonumber\\
&-& \frac{(\alpha - r)^2 Y_z^2(z) - 2 (\alpha - r) \eta \rho Y_z (z) (Y_z (z) + z Y_{zz} (z)) + \eta^2 \rho^2 \sigma^2 (Y_z (z) - z Y_{zz} (z))^2}{2 \sigma^2 Y_{zz} (z)} \nonumber \\
&-&   \log Y_z (z) + \log \kappa  - 1= 0.
\end{eqnarray}
Naturally, the investment $\pi(z, \tau ,h)$ and consumption $c(z, \tau , h)$ in terms of $Y(z)$ look now as follows

\begin{eqnarray}
		\pi (z, \tau ,h) &=& h \left( \frac{\eta \rho}{\sigma}z + \frac{\eta \rho \sigma - \alpha + r}{\sigma^2} \frac{Y_z (z)}{Y_{zz}(z)} \right), \label{pi:LOGexpH8} \\
		 c(z, \tau ,h) &=& h \frac{\kappa^2}{Y_z (z)}. \label{c:LOGexpH8}
\end{eqnarray}
In terms of original variables $l, h, t$ and $V$ the substitution looks as follows
\begin{eqnarray}
z &=& \frac{l}{h}, ~~~ \tau =  rt - \log h, \label{2subHARAH8}  \\
Y(z) &=&  \kappa e^{\kappa t} V  - \log h. \nonumber
\end{eqnarray}
It also means that if we obtain a solution $Y(z)$ for (\ref{eq:HARAH8}) we obtain the value function that in terms of original variables looks like
$$
V(t, l ,h) =   \frac{e^{- \kappa t}}{\kappa} Y(l/h) +  \frac{e^{- \kappa t}}{\kappa} \log h,
$$
and the condition $V(l, h, t) \underset{t \to \infty}{\longrightarrow} 0$ is satisfied.

All other two dimensional subalgebras listed in Table \ref{TableOptSystL4LOG} do not give a reduction of the original equation (\ref{eq:LOGexp}) to an ODE, so we will not regard them in detail.

\section{Conclusion}

In this paper we regard a portfolio optimization problem for a basket consisting of a riskless liquid, risky liquid and risky illiquid assets. The illiquid asset is sold in a random moment of time $T$ that has a distribution with a survival function $\overline{\Phi} (t)$, satisfying very general conditions  $\lim_{t \to \infty} \overline{\Phi} (t) E[U(c(t))]=0$ and $\overline{\Phi} (t) \sim e^{-\kappa t}$ or faster as $t \to \infty$. This, to our knowledge, is a new generalization of the illiquidity models that was for the first time introduced in \cite{boyazh}.

 We work with two different utility functions (i.e.  logarithmic and HARA-utility). Both of the utility functions were widely used before for the problems of random income and for the portfolio optimization with a portfolio that includes an illiquid asset sold in a deterministic moment of time or with infinite time horizon. In our setting we introduce an exogenous random liquidation time, i.e. a random moment of time T such that the illiquid asset in the optimized portfolio is sold (i.e. liquidated) at this moment. We use two types of utility functions: HARA type utility and logarithmic utility. However, we demonstrate that there is a possibility to choose HARA-utility in such a form that  $U^{HARA}\underset{\gamma \to 0}{\longrightarrow} U^{LOG}$. This fact was mentioned in some publications before but, to our knowledge, it was not demonstrated explicitly, in a step-by-step manner how deep this connection is on the analytical level. May be because of that majority of the HARA utilities that we found in the literature do not actually posses this quality. They are certainly  utility functions of the HARA type but the limit procedure applied to such utility function can at its best be reduced to some modification of logarithmic utility (not $\log c$), while some of them do not even have anything that remotely resembles a logarithm. To demonstrate the connection between two problems we choose $U^{HARA}$ in the form (\ref{harautility}) and show that not only we obtain correct form of logarithmic utility as $\gamma \to 0$, i.e. $U^{HARA}\underset{\gamma \to 0}{\longrightarrow} U^{LOG}$ but, naturally, a three-dimensional HJB equation (\ref{eq:HJB21}) corresponding to HARA-utility formally transforms into an HJB-equation that corresponds to logarithmic case as $\gamma \to 0$. After a formal maximization of (\ref{eq:HJB21}) we obtain three dimensional PDEs corresponding to HARA and logarithmic utility correspondingly. We demonstrate that the PDE (\ref{maingeneralLOG})  arising in the case of logarithmic utility function can be formally regarded as a limit case of the PDE (\ref{maingeneralHARA}) arising in the case of HARA utility function as $\gamma \to 0$ . 
 
 We carry out the Lie group analysis of the both three dimensional PDEs and are able to obtain the admitted symmetry algebras. Then we prove that the algebraic structure of the PDE with logarithmic utility can be seen as a limit of the algebraic structure of the PDE with HARA-utility as $\gamma \to 0$. Moreover, this relation does not depend on the form of the survival function $\overline{\Phi} (t)$ of the random liquidation time $T$. Therefore, we demonstrate that if HARA utility has a special form such that $U^{HARA}\underset{\gamma \to 0}{\longrightarrow} U^{LOG} $, formally $ L^{HARA}_{3,4} \underset{\gamma \to 0}{\longrightarrow} L^{LOG}_{3,4}$ and a logarithmic utility can be regarded as a limit case of HARA utility and this correspondence holds not only for the HJBs but also for the algebraic structures, that stand behind them.

We carry out a complete Lie-symmetry analysis for two different utility functions, i.e. for two different three dimensional PDEs (\ref{maingeneralHARA}) and (\ref{maingeneralLOG}) which contain an arbitrary function $\overline{\Phi} (t)$. In both cases we are able to solve these rather voluminous problems and find the admitted Lie algebras $L^{HARA}_3$ and $L^{LOG}_3$. The study of the three dimensional problems is connected with a lot of tedious calculations, even the first step on which one needs to find the determining system is, in fact, non-trivial. These difficulties become even more evident if we have an arbitrary function in the studied equation. The problem becomes slightly more tractable if one applies package {\bf IntroToSymmetry}, but the majority of the calculations still are to be done manually. In this way we get the system of partial differential equations containing 137 and 130 different equations correspondingly. These equations define the generators of the corresponding algebras. Computer systems that we know of, unfortunately, can not solve the obtained system of differential equations. This has to be a step-by-step handmade procedure. To our knowledge, this is a first Lie group analysis of such problem for a general liquidation time distribution.  If we look on the amount of calculations it is also understandable why just few cases of three dimensional PDEs are profoundly studied in the literature.

We study the internal structure of the admitted algebras further in order to use these structures and obtain convenient and useful reductions of both PDEs (\ref{maingeneralHARA}) and (\ref{maingeneralLOG}). Using the notation provided in \cite{patera&wintern} we show that for a general liquidation time distribution $L^{HARA}_3$ can be classified  as $A^{\gamma}_{3,5}$ and $L^{LOG}_3$ as $A_1 \bigoplus A_2$. We use the system of optimal subalgebras provided in \cite{patera&wintern} and obtain corresponding reductions of both three dimensional PDEs. We show how each of the original problems can be reduced to a corresponding two dimensional one and prove that in general case with an arbitrary function $\overline{\Phi} (t)$ there is no Lie type reduction that leads to an ODE.

We also show that if and only if the liquidation time defined by a survival function $\overline{\Phi} (t)$  is distributed exponentially, then for both types of the utility functions we get an additional symmetry. We prove that both Lie algebras admit this extension, i.e. we obtain the four dimensional $L^{HARA}_4$ and $L^{LOG}_4$ correspondingly for the case of exponentially distributed liquidation time.
Indeed, the case of exponentially distributed liquidation time is actually similar to the infinite-horizon random income problem and several other models studied in the literature (see, for instance, \cite{ZaripDufFlem}), yet our work is the first to our knowledge that explicitly shows which properties make an exponentially distributed liquidation time a distinguished case, that allows a reduction of an original three-dimensional PDEs to ODEs. This is a very important result, since a lot of the works in the field use similar reductions and do not mention that there is actually no other distribution that allows a Lie type reduction of a PDE to an ODE.
With a help of Lie group analysis we explain what makes exponential liquidation time distribution a distinguished case, the only situation when one can reduce the three dimensional HJBs to ODEs.

We show that for both utility functions the symmetry algebras $L^{HARA}_4$ and $L^{LOG}_4$ admitted by the corresponding equations (\ref{eq:HARAexp}) and (\ref{eq:LOGexp}) can be classified as $A_2 \oplus A_2$ in terms of the notation used in \cite{patera&wintern}.
The study of the Lie algebraic structure of both three dimensional PDEs gives rise to an interesting remark on different level of influence associated with different parameters that define the model. Both equations contain a lot of parameters which describe properties of assets, utility functions and liquidation time distributions. From all of these parameters just three parameters have deep influence on the algebraic structure of the admitted Lie algebras: in both cases it is the interest rate $r$, the parameter $\kappa$ of the exponential time distribution and for HARA utility function it is its' parameter $\gamma$. The algebras change their structure if one or some of these parameters vanishing.
Additionally if $\kappa = r \gamma$, i.e. there is a certain dependency between the parameter of HARA utility function and the corresponding liquidation time distribution,  the symmetry algebra $L^{HARA}_4$ has the structure that corresponds to $A^{\gamma}_{3,5} \bigoplus A_1$, in the notation of \cite{patera&wintern}.

Using a system of optimal subalgebras for all admitted algebras allows us to provide all non equivalent reductions of the studied equations and describe the solutions which can not be transformed to each other with a help of the transformations from the admitted symmetry group. For every case we list all possible Lie type reductions of the problem. The reduced equations that are two dimensional PDEs or in some special cases are even ODEs. Such equations are much more convenient for further analytical or numerical studies.

We also show how one can rewrite corresponding optimal policies in every case. They can be described using solutions of the reduced equations.
One can see that analogously to the result, obtained in \cite{BordagYamshchikovZhelezov} optimal policies tend to classical Merton policies as $h \to 0$, that is only logical, since by construction this situation should correspond to a portfolio without illiquid asset.

 Summing up, we carry a complete Lie group analysis for two different optimization problems with HARA and logarithmic utility functions and for a general as well as exponential liquidation time distributions. We list reduced equations and corresponding optimal policies that tend to the classical Merton policies as illiquidity becomes small.

\begin{acknowledgements}
The authors are thankful to Prof. F. Oliveri who provided introduction and support by using his software packages {\bf ReLie} and {\bf SymboLie}.\\
This research was supported by the European Union in the FP7-PEOPLE-2012-ITN Program under Grant Agreement Number 304617 (FP7 Marie Curie Action, Project Multi-ITN STRIKE - Novel Methods in Computational Finance)
\end{acknowledgements}



% Author, Article title, Journal, Volume, page numbers (year)
% Author, Book title, page numbers. Publisher, place (year)

\bibliographystyle{plain}
\addcontentsline{toc}{chapter}{Bibliography}
\bibliography{[]Collection_for_book}

\end{document}